\documentclass{emulateapj}
\usepackage{apjfonts,color,xspace}

\usepackage{ifthen}
\newcommand{\zone}[1]{%
\ifthenelse{\equal{69}{#1}}{A}{}%
\ifthenelse{\equal{209}{#1}}{B}{}%
\ifthenelse{\equal{556}{#1}}{C}{}%
\xspace}
\newcommand{\Zone}[1]{Zone~\zone{#1}}
\newcommand{\Msun}{\ensuremath{\mathrm{M}_\odot}}

\begin{document}

\title{The interplay between chemistry and nucleation in the formation
  of carbonaceous dust in supernova ejecta}

\shorttitle{Carbonaceous dust from CCSNe}

\author{Davide Lazzati\altaffilmark{1}, Alexander
  Heger\altaffilmark{2,3,4,5}}

\shortauthors{Lazzati \& Heger}

\altaffiltext{1}{Department of Physics, Oregon State University, 301
Weniger Hall, Corvallis, OR 97331, U.S.A.}
\altaffiltext{2}{Monash Center for Astrophysics, School of
  Physics and Astronomy, Monash University, Vic 3800, Australia}
\altaffiltext{3}{School of Physics \& Astronomy,
  University of Minnesota, Minneapolis, MN 55455, U.S.A.}
\altaffiltext{4}{Joint Institute for Nuclear Astrophysics, 225
  Nieuwland Science Hall Department of Physics, University of Notre
  Dame, Notre Dame, IN 46556, U.S.A.}
\altaffiltext{5}{Center for Nuclear Astrophysics, Department of
  Physics and Astronomy, Shanghai Jiao-Tong University, Shanghai
  200240, P. R. China.}

\begin{abstract}
  Core-collapse supernovae are considered to be important contributors
  to the primitive dust enrichment of the interstellar medium in the
  high-redshift universe.  Theoretical models of dust formation in
  stellar explosions have so far provided controversial results and a
  generally poor fit to the observations of dust formation in local
  supernovae.  We present a new methodology for the calculation of
  carbonaceous dust formation in young supernova remnants.  Our new
  technique uses both the nucleation theory and a chemical reaction
  network to allow us to compute the dust growth beyond the molecular
  level as well as to consider chemical erosion of the forming grains.
  We find that carbonaceous dust forms efficiently in the core of the
  ejecta, but takes several years to condensate, longer than
  previously estimated.  It forms unevenly and remains concentrated in
  the inner part of the remnant.  These results support the role of
  core-collapse supernovae as dust factories and provide new insight
  on the observations of SN~1987A, in which large amounts of dust have
  been detected to form on a timescale of years after core collapse.
\end{abstract}


\section{Introduction}

The observations of significant amounts of dust in high redshift
quasars (Bertoldi et al.\ 2003; Priddey et al.\ 2003; Robson et
al.\ 2004; Beelen et al.\ 2006; Wang et al.\ 2008) have highlighted the
requirement for an alternative to asymptotic giant branch (AGB) stars
as the dominant dust factories in the early Universe (e.g.,
Maiolino et al.\ 2004, Valiante et al.\ 2009, 2011).  An ideal
alternative are core-collapse supernovae (CCSNe), since the short main
sequence time of their progenitor stars allows them to enrich the
interstellar medium with the first solids with no noticeable delay on
cosmological timescales.

Theoretical calculations of the dust yields of CCSNe, however, have
produced controversial results.  Calculations based on the classical
nucleation theory (e.g., Feder et al.\ 1966) typically predict the
prompt formation of fairly large amounts of dust, of the order of one
solar mass or more, within one year from the explosion (Kozasa et al.\
1989, 1991; Clayton et al.\ 2001; Todini \& Ferrara 2001; Nozawa et
al.\ 2003, 2010).  The result is robust to the unknown parameters of
the theory, such as the shape and sticking coefficients of the grains,
at least to first order (Fallest et al.\ 2013).  On the other hand,
calculations that model the formation of solid grains with a chemical
network with well-defined reaction rates resulted in the prediction of
only modest amount of dust, its condensation starting only years after
the supernova explosion (Cherchneff \& Dwek 2009, 2010; Sarangi \&
Cherchneff 2013; Biscaro \& Cherchneff 2014).  Such calculations,
however, cannot follow the grain formation up to large molecular
clusters and/or micron-sized crystals. Most current results only
predict the formation of molecular precursors, such as C$_{12}$. A
recent paper (Sarangi et al. 2015) follows carbon cluster formation up
to C$_{28}$, assuming that coalescence is responsible for the growth
of astrophysical grains from the clusters.

The discrepancy between the two theoretical predictions seems to be
brought about by the approximations and intrinsic limitations of the
two theories.  Calculations based on the nucleation theory assume that
the grain formation happens in virtual isolation, ignoring the fact
that the growing proto-grains are exposed to the rich field of
damaging radiation of a young supernova remnant (SNR).  They also
assume that even the smallest molecular clusters are continuum
entities, ignoring their atomic structure and quantum behaviour.  On
the other hand, chemical network calculations provide a deterministic
growth of grains and therefore ignore the fact that, as grains grow
beyond the molecular cluster phase, self shielding allow the few
grains that grew by statistical fluctuations to become more and more
resilient and grow even further.  Chemical calculations, as such,
ignore the fact that grain growth becomes a runaway process, the few
grains able to make it beyond a few hundred monomers allowed to grow
and become stable.

Not surprisingly, observations of nearby core-collapse supernovae
(SN1987A and, only recently, SN2010jl) are in disagreement with both
results.  They show dust formation beginning as early as half a year
after the explosion (Gall et al. 2014; earlier dust formation being
possible but hidden under the more abundant curcumstellar dust
component), in disagreement with chemical network results.  The early
dust formation, however, is modest, at most a fraction of a per cent
of a solar mass being condensed (in disagreement with the classical
nucleation results).  To add confusion to the picture, recent Herschel
(Matsuura et al.\ 2013) and ALMA (Indebetouw et al.\ 2014)
observations of SN1987A show the presence of large amounts of dust in
emission in the center of the remnant.  Taken at face value, the
collective data from all nearby core-collapse supernovae suggests a
scenario in which dust formation is a slow and continuous process that
starts at most a few months after core collapse and continues for at
least several years, eventually yielding amounts of solid particles
totalling approximately one solar mass. Not all core-collapse
supernovae may be dust producers, however. In a recent study, Szalai
\& Vink\'o (2013) find that only two out of twelve type II-P SNe show
signs of newly sinthsized dust in emission.

Connecting the chemistry of the molecular precursor with a formalism
for dust growth is necessary to properly describe the formation and
growth of grains into stable microscopic solids (Sarangi \& Cherchneff
2015). In this paper we present a new framework for the calculation of
dust formation in supernova explosions based on the kinetic nucleation
theory (e.g., Kashchiev 2000).  Whereas the kinetic nucleation theory
is equivalent to the standard thermodynamic theory for homologous
nucleation in isolation, it allows to introduce deviations from the
capillary approximation for very small clusters and chemical
weathering of the forming grains in the harsh supernova environment.
The theory is presented, and applied as a proof of concept, to the
formation of carbonaceous grains, but can be readily extended to the
formation of any grain species, provided that its susceptibility to
weathering in the remnant environment is known.

This paper is organized as follows: in Section~2 we present the
chemical network and the nucleation theory that we adopt.  In
Section~3 we describe the stellar explosions model adopted and in
Section~4 we present out results.  A summary and discussion are in
Section~5.

\section{Chemistry and dust formation physics}

This work is based on a small chemical and dust nucleation network
aimed at keeping track of the concentrations of CO molecules, free
oxygen and carbon atoms, hydrogen and helium cations, thermal and
non-thermal free electrons, and carbonaceous grains. In this section
we describe the physical processes and the techniques used to compute
such concentrations. The reactions considered are:
\begin{eqnarray}
\mathrm{He} + \mathrm{e}^-_\mathrm{C}  \;&\to& \; \mathrm{He}^+ + \mathrm{e}^-_\mathrm{C} + \mathrm{e}^-_\mathrm{T} \label{reac:1}\\
\mathrm{He}^+ + \mathrm{e}^-_\mathrm{T}  \;&\to& \; \mathrm{He} \label{reac:2}\\
\mathrm{H} + \mathrm{e}^-_\mathrm{C}  \;&\to& \; \mathrm{H}^+ + \mathrm{e}^-_\mathrm{C} + \mathrm{e}^-_\mathrm{T} \label{reac:3}\\
\mathrm{H}^+ + \mathrm{e}^-_\mathrm{T}  \;&\to& \; \mathrm{H} \label{reac:4}\\
\mathrm{C}+\mathrm{O}  \;&\to& \; \mathrm{C}\mathrm{O} \label{reac:5}\\
\mathrm{C}\mathrm{O}  \;&\to  \;& \mathrm{C}+\mathrm{O} \label{reac:6}\\
\mathrm{C}\mathrm{O}+ \mathrm{e}^-_\mathrm{C}  \;&\to& \; \mathrm{C}+\mathrm{O}+ \mathrm{e}^-_\mathrm{C} \label{reac:7}\\
\mathrm{C}\mathrm{O}+ \mathrm{He}^+  \;&\to& \; \mathrm{O}+\mathrm{He}+\mathrm{C}^+ \label{reac:7bis}\\
\mathrm{C}_n+\mathrm{O}  \;&\to& \; \mathrm{C}_{n-1}+\mathrm{C}\mathrm{O} \label{reac:O}\\
\mathrm{C}_n+\mathrm{He}^+  \;&\to  \;& \mathrm{C} _{n-1}+\mathrm{He}+\mathrm{C}^+ \label{reac:He}\\
\mathrm{C}_n+\mathrm{C}  \;&\to& \; \mathrm{C} _{n+1}\label{reac:10}\\
\mathrm{C}_n \;&\to& \; \mathrm{C} _{n-1}+\mathrm{C} \label{reac:11}
\end{eqnarray}
where $\mathrm{e}^-_\mathrm{C}$ is a Compton electron consequence of
the energy deposited by the decay of $^{56}$Ni into $^{56}$Co and
eventually $^{56}$Fe and $\mathrm{e}^-_\mathrm{T}$ is a free electron
in thermal equilibrium with the surrounding gas.

Reactions~\ref{reac:1} and~\ref{reac:2} are the ionization and
recombination of He atoms as a consequence of the presence of
non-thermal electrons.  Reactions~\ref{reac:3} and~\ref{reac:4} are
the ionization and recombination of H atoms as a consequence of the
presence of non-thermal electron.  Reaction~\ref{reac:5} is the
formation of carbon monoxide molecules by radiative association, and
reactions~\ref{reac:6}, \ref{reac:7}, and~\ref{reac:7bis} are the
destruction of the CO molecule due to thermal effects, interaction
with non-thermal electrons, and charge-exchange with He$^+$ cations,
respectively.  Reactions~\ref{reac:O} and~\ref{reac:He} are the
chemical weathering reactions in which the Carbon clusters or grains
lose one carbon atom by reaction with a free oxygen atom (oxidation)
or He ion (ion-molecule).  Finally, reactions~\ref{reac:10}
and~\ref{reac:11} are the monomer attachment and spontaneous
detachment reactions to and from carbon clusters.  The network
considered here is much smaller than the one used by Cherchneff \&
Dwek (2009, 2010).  In this paper, however, we are focusing on the
carbonaceous nucleation only rather than following all possible dust
species.  In addition, the main goal of this work is to study the
interplay between chemistry and nucleation with an hybrid approach
that allows us to consider carbon clusters both as molecules, subject
to interactions with other species, and as small solids, capable of
growing indefinitely.

\subsection{Ionization balance}

To compute the density of free electrons and of the HII and HeII
cations, we calculate the balance between the ionization rate due to
interaction with Compton electrons and recombination from the bath of
thermal electrons.  We then use the H and He ionization as a proxy for
the ionization of the whole SNR to compute the total density of free
electrons.

The ionization rate of the atom or ion $X$ due to collisions with
Compton electrons is given by:
\begin{equation}
k_X=L_\gamma/W_X\;,
\label{eq:ion1}
\end{equation}
where $W_X$ is the average energy per ionization of the species $X$
(Liu \& Victor 1994; Liu \& Dalgarno 1995; Cherchneff \& Lilly
2008). The luminosity of the $^{56}$Ni decay $L_\gamma$ is given by
(e.g., Woosley et al.\ 1989):
\begin{equation}
L_\gamma=7.5\times10^{-8}
\frac{N_{^{56}\mathrm{Ni}}}{N_{\rm{tot}}}
\langle{E_\gamma}\rangle
f_\gamma(K_{56})
e^{-\frac{t}{\tau_{56}}}
\quad
\mathrm{Me\!V}
\;
\rm{s}^{-1}
\label{eq:ion2}
\end{equation}
where $N_{^{56}\mathrm{Ni}}$ is the total number of $^{56}$Ni atoms
produced in the SN explosion, $N_{\rm{tot}}$ is the total number of
atoms in the ejecta, $\langle{E_\gamma}\rangle=3.57\,\mathrm{Me\!V}$
is the average energy released in each decay, and
$\tau_{56}=111.26\,$d is the $e$-folding time of $^{56}$Co
decay. $f_\gamma(K_{56})$, the deposition function, is given by:
\begin{equation}
f_\gamma(K_{56})=1-e^{-K_{56}\phi_0(t_0/t)^2}\;,
\label{eq:ion3}
\end{equation}
where $K_{56}=0.033$ is the average $\gamma$-ray mass absorption
coefficient (Woosley et al.\ 1989), and
$\phi_0=5\times10^{4}\,\mathrm{g}\,\mathrm{cm}^{-2}$ is the mass
column density of the ejecta model used in this work at a reference
time $t_0=10^6\,$s (Woosley et al.\ 1989; Todini \& Ferrara 2001).
Most of the numerical values used were computed for models of the
SN~1987A explosion and may not be strictly appropriate for our
computation.  Given the overall simplification of the model, however,
and especially the assumption of a uniform density of non-thermal
electrons throughout the remnant, computing appropriate values for the
SN explosion considered here is beyond the scope of the paper.  The
average energy per ionization of H is (Dalgarno et al.\ 1999)
\begin{equation}
W_{\rm{H}}=36.1 \, \rm{e\!V}
\end{equation}
and the average energy per ionization of He is
\begin{equation}
W_{\rm{He}}=46.3 \, \rm{e\!V} \;.
\end{equation}
The temperature-dependent recombination rate $\alpha_\mathrm{r}(T)$ of
free thermal electrons on the hydrogen and helium ions was computed
following Verner \& Ferland (1996).  The recombination rate is given
(in units of cm$^{3}$~s$^{-1}$) by:
\begin{equation}
\alpha_\mathrm{r}(T)=a\left[\sqrt{\frac{T}{T_0}}\left(1+\sqrt{\frac{T}{T_0}}\right)^{1-b}
  \left(1+\sqrt{\frac{T}{T_1}}\right)^{1+b}\right]^{-1}\;.
\label{eq:rec}
\end{equation}
The parameters $a$, $b$, $T_0$, and $T_1$ are given in
Table~\ref{tab:rec}.

\begin{table}
\begin{tabular}{c|c|c|c|c}
& $a$ (cm$^3$~s$^{-1}$)& $b$ & $T_0$ & $T_1$ \\ \hline \hline
H & $7.982\times10^{-11}$ & 0.748 & 3.148 & $7.036\times10^5$ \\
He & $9.356\times10^{-10}$ & 0.7892 & $4.266\times10^{-2}$ &
$4.677\times10^6$
\end{tabular}
\caption{{Parameters used in Eq.~\ref{eq:rec} to compute the
  recombination rates of hydrogen and helium cations.}
\label{tab:rec}}
\end{table}

\subsection{CO Molecule}
\label{sec:CO}

For the carbon monoxide molecule, we consider one formation process
(radiative association) and three destruction processes: thermal
destruction due to interactions with the surrounding gas in
thermodynamic equilibrium, non-thermal destruction due to collisions
with Compton electrons, and destruction due to the reaction with a
He$^+$ cation.  The formation rate was computed using an analytic
interpolation of the tabulated data of Dalgarno et al.\ (1990):
\begin{equation}
^+k_\mathrm{CO}=\frac{4.467\times10^{-17}}
  {\sqrt{\left(\frac{T}{4467}\right)^{-2.08}+\left(\frac{T}{4467}\right)^{-0.22}}}
\end{equation}

In the absence of non-thermal electrons, the concentration of the CO
molecule is given by (e.g., Clayton 2013):
\begin{equation}
n_\mathrm{CO}=n_\mathrm{C} n_\mathrm{O}\left(\frac{h^2}{2\pi{}MkT}\right)^{3/2}e^\frac{B_\mathrm{CO}}{kT}
\end{equation}
where $h$ is Planck's constant, $k$ is Boltzmann's constant, $M$ is
the reduced mass of C and O, and $B_\mathrm{CO}=11.1\,\mathrm{e\!V}$
is the binding energy of the CO molecule.  At equilibrium,
\begin{equation}
\frac{dn_\mathrm{CO}}{dt}=0=\,^+k_\mathrm{CO}n_\mathrm{C} n_\mathrm{O}-^-k_\mathrm{CO} n_\mathrm{CO}
\end{equation}
where $^-k_\mathrm{CO}$ is the thermal destruction rate of CO.  This
yields
\begin{equation}
^-k_\mathrm{CO}=\frac{^+k_\mathrm{CO}n_\mathrm{C}n_\mathrm{O}}{n_\mathrm{CO}}=
\,  ^+k_\mathrm{CO}\left(\frac{h^2}{2\pi{}MkT}\right)^{-3/2}e^{-\frac{B_\mathrm{CO}}{kT}}\;.
\end{equation}
Note that we do not compute the CO abundance at equilibrium. We only
assume, throughout this calculation, that the thermal destruction
coefficient of the CO molecule is independent on the density of the
molecule and therefore the destruction coefficient at equilibrium is
the same as the coefficient out of equilibrium. A somewhat different
value was obtained experimentally by Appleton et al.\ (1970), and used
by Cherchneff \& Dwek (2009, 2010). The Appleton et al.\ (1970)
experiment, however, was carried out at much higher temperature
($8000-15000$~K) and may not be precise in the temperature range
considered here.

The rate of CO molecule destruction due to interaction with
non-thermal electron was computed with Equations~\ref{eq:ion1},
\ref{eq:ion2}, and~\ref{eq:ion3} adopting (Liu \& Dalgarno 1995):
\begin{equation}
W_{\mathrm{CO}}=125 \, \mathrm{e\!V}
\end{equation}

Finally, the CO destruction rate by interactions with He$^+$ cations
was computed from the Arrhenius coefficient reported in the Umist
Database for Astrochemistry 2012 (McElroy et al. 2013;
{\tt{www.udfa.net}}):
\begin{equation}
^-k_{\mathrm{CO,He^+}}=1.6\times10^{-9}
\end{equation}

\subsection{Dust nucleation and growth}

The nucleation of stable carbonaceous grains is modeled following the
kinetic theory of dust nucleation (Becker \& D\"oring 1935; Kashchiev
2000), assuming local-thermal equilibrium and adopting the capillary
approximation for clusters with more than two carbon atoms.  The
kinetic theory of dust nucleation describes the nucleation process as
the result of a competition between attachment and detachment of
carbon atoms to and from existing, unstable molecular clusters.  The
nucleation rate is given by the master equation,
\begin{equation}
J=n_{C,gas}f_1\left[1+\sum_{i=2}^\infty\left(\prod_{j=2}^i
\frac{g_i}{f_i}\right)\right]^{-1}\;,
\label{eq:J}
\end{equation}
in which $n_{\mathrm{C,gas}}$ is the number density of carbon in the gas
phase, $f_1$ is the rate of the
$\mathrm{C}+\mathrm{C}\to\mathrm{C}_2$ formation reaction, $f_i$ is
the rate of the $\mathrm{C}_i+\mathrm{C}\to\mathrm{C}_{i+1}$ growth
process, and $g_i$ is the rate of the
$\mathrm{C}_i\to\mathrm{C}_{i-1}+\mathrm{C}$ detachment process.

The growth rate of the $i$-sized cluster, $f_i$, is taken to be the
fraction $\lambda$ of the impingement rate of gas phase carbon atom
onto the cluster.  The impingement rate can be derived from the kinetic
theory of gases yielding a growth rate:
\begin{equation}
f_i=\lambda \, n_\mathrm{C,gas} \, c_\mathrm{s} \, \left(i\, v_\mathrm{C}\right)^{2/3}
\sqrt{\frac{kT}{2\pi m_\mathrm{C}}}
\label{eq:fi}
\end{equation}
where $v_\mathrm{C}=8.933\times10^{-24}\,\mathrm{cm}^3$ is the volume
occupied by a carbon molecule in the solid phase,
$m_\mathrm{C}=2\times10^{-23}\,\mathrm{g}$ is the mass of a carbon
atom, and $c_\mathrm{s}=S/V^{2/3}$ the shape factor ($S$ being the
grain surface and $V$ its volume).  Analogously to previous work on
dust formation in core-collapse supernova explosions, we assume
throughout this paper $\lambda=1$ and $c_\mathrm{s}=(36\pi)^{1/3}$,
the shape factor of spherical grains.  For a test on how assuming
different values for these parameters affects the amount and timing of
dust condensation, see Fallest et al.\ (2013).

\begin{figure}
\includegraphics[width=\columnwidth]{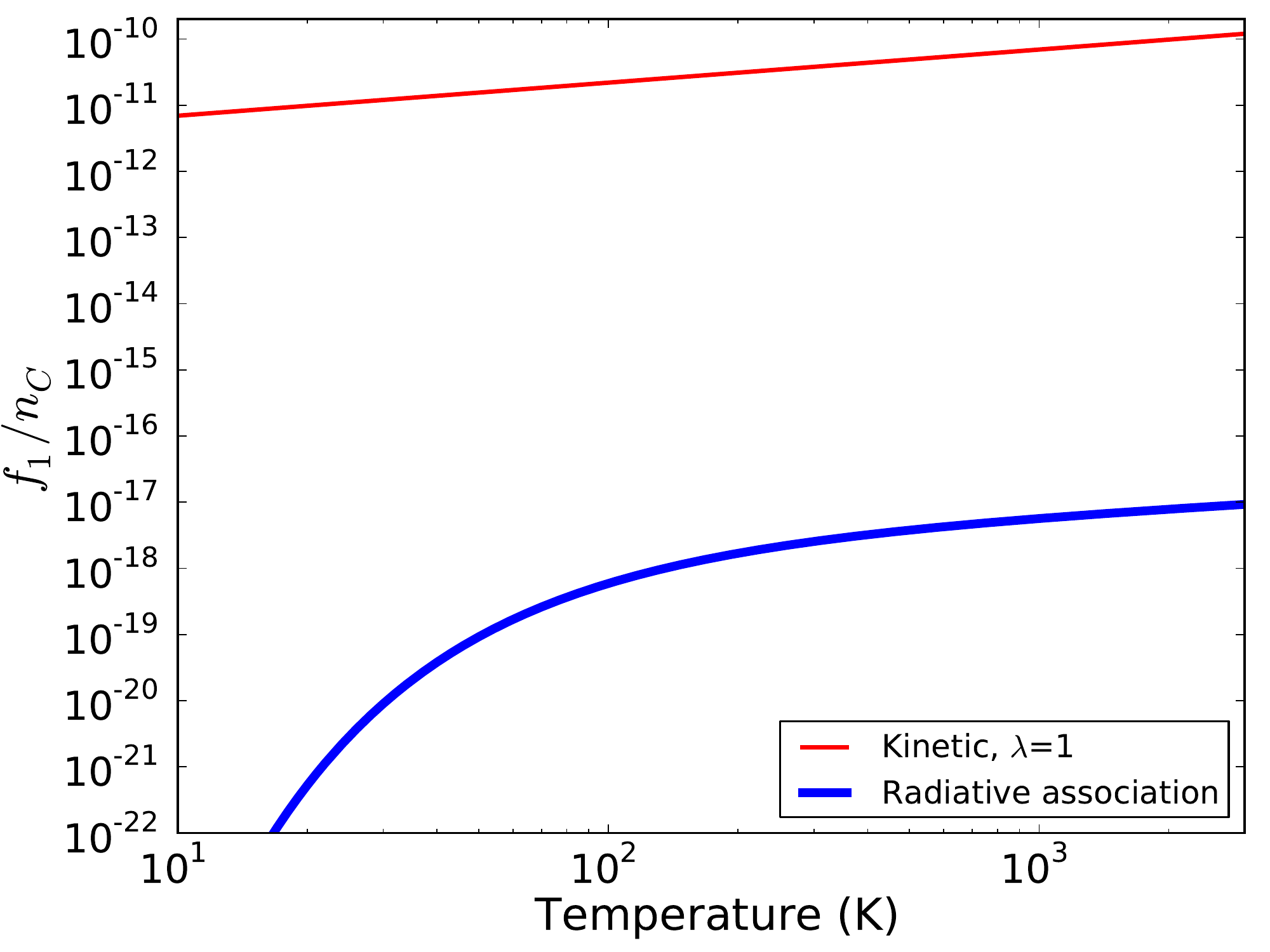}
\caption{Comparison between the density normalized attachment rate of
    C atoms on C atoms to form a dicarbon molecule. The thick blue
    line shows the radiative association rate (from Andreazza and
    Singh 1997) that is used in this work. The thin red line shows
    the limit $i=1$ of Eq.~\ref{eq:fi}, implicitly adopted in most
    dust nucleation studies.
\label{fig:f1}}
\end{figure}

\begin{figure}
\includegraphics[width=\columnwidth]{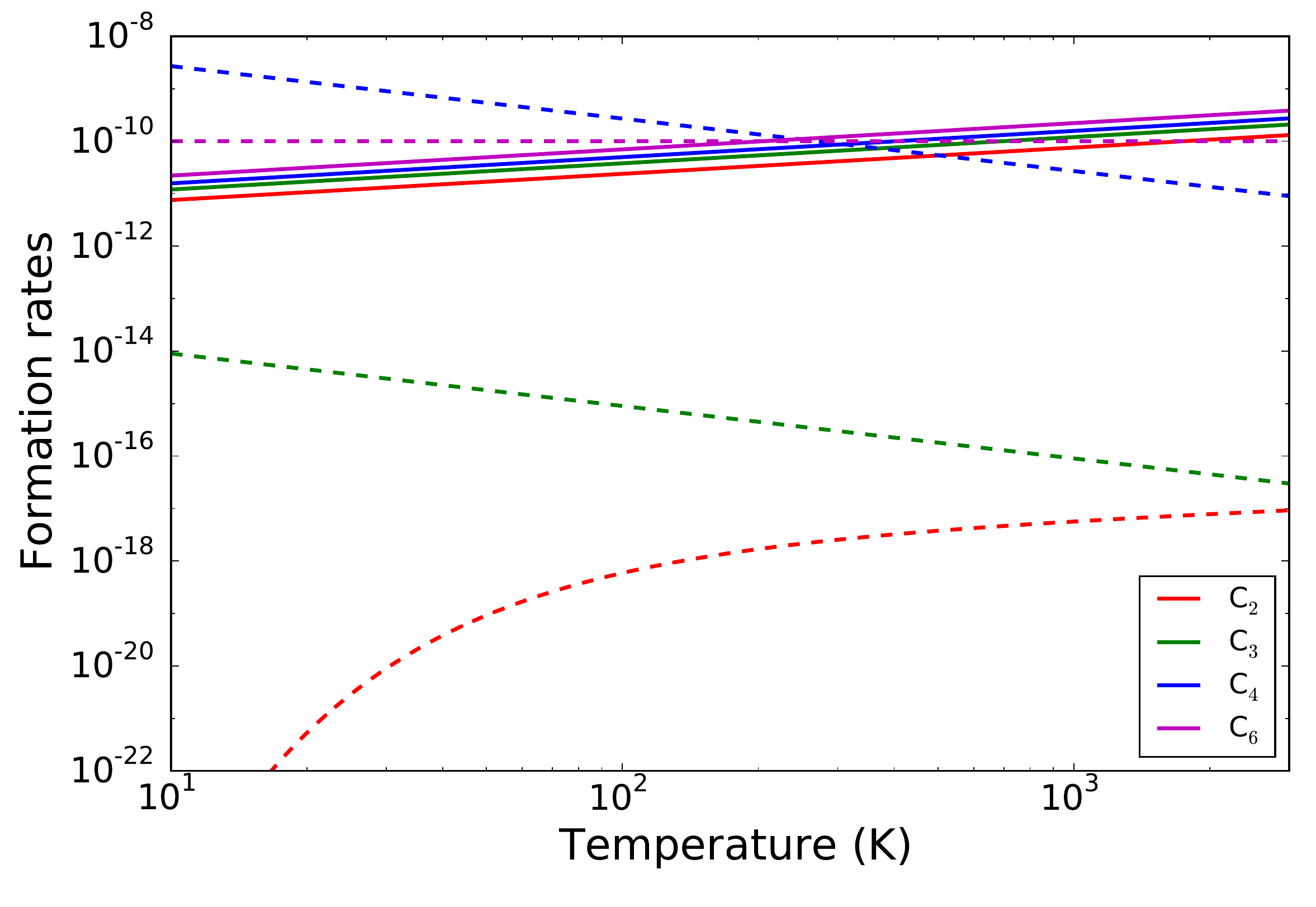}
\caption{{Comparison between the formation rate of carbon clusters of
    different size by monomer accretion as computed by radiative
    association (dashed lines) and by the nucleation theory (solid
    lines). In this paper we adopt the rates from the nucleation
    theory for any cluster with more than two atoms. See the text for
    a discussion.}
\label{fig:rates}}
\end{figure}

A dedicated discussion is required for the formation of a dicarbon
molecule from two C atoms. When a C atom approaches a $\mathrm{C}_i$
carbon cluster and binds to it, the excess energy is dissipated
through an increase in the cluster temperature, distributed among its
$3i-6$ vibrational degrees of freedom.  This allows for non-radiative
associations with a sticking coefficient of order unity.  The same,
however, cannot be said for the dimer formation.  A dimer only has one
vibrational degree of freedom and, should all the excess energy be
dumped there, it would be large enough to break the bond and return
the two $C$ molecules to the gas phase.  For this reason, instead of
taking the limit $i=1$ of Eq.~\ref{eq:fi}, we adopt the $\mathrm{C}_2$
formation rate from radiative association (Andreazza \& Singh 1997):
\begin{equation}
f_1=4.36\times10^{-18}\left(\frac{T}{300}\right)^{0.35}
e^{-\frac{161.31}{T}}n_\mathrm{C,gas}
\label{eq:C2}
\end{equation}
A comparison between this rate and the rate obtained by using $i=1$ in
Eq.~\ref{eq:fi} is shown in Figure~\ref{fig:f1} as a function of
temperature.  The striking difference of many orders of magnitude
between the two curves is a warning flag of how inadequate are some
assumptions currently made in dust condensation
calculations\footnote{An alternative path to the carbon dimer involves
  the formation of a CO molecule and its reaction with a free carbon
  atom. This was shown to be less effective by Yu et al.\ (2013).}
(see also Donn \& Nuth 1985). What happens for the formation of the
C$_i$ molecules with $i=3,4,5$, and up to a few tens is not fully
understood. Rates for radiative association have been computed by
Wakelam et al.\ (2009) for a few carbon clusters (C$_3$, C$_4$, C$_6$,
and C$_8$). Their results are compared to the nucleation theory rates
in Figure~\ref{fig:rates}. The difference between the nucleation and
radiative association rates is, again, quite large. For the formation
of clusters with more than two atoms, the reaction can be radiative or
non-radiative, as assumed in the nucleation theory, since the extra
energy of the incoming monomer can be absorbed in the 3i-6 degrees of
freedom of the accreting cluster.  Given the lack of a robust and
complete set of radiative reaction rates, we here perform our
calculations with the rates from the nucleation theory, the solid
lines in Figure~\ref{fig:rates}, as derived from Eq.~\ref{eq:fi}.

In the absence of harmless radiation and cations, the ejection rates
$g_i$ can be computed by enforcing detailed equilibrium in a canonical
ensemble of carbon particles at saturation.  In such an ensemble, true
equilibrium can be achieved, and the distribution of the number of
clusters of size $i$ is given by the Boltzmann equation,
\begin{equation}
n_i\propto e^{\frac{\sigma_\mathrm{C} S_i}{kT}}\;,
\label{eq:ni}
\end{equation}
where $\sigma_\mathrm{C}$ is the surface energy of solid phase carbon
and $S_i$ the surface area of the $i$-sized cluster.  Detailed
equilibrium reads:
\begin{equation}
n_i g_i = n_{i-1} f_{i-1}
\label{eq:deteq}
\end{equation}
from which we find
\begin{eqnarray}
g_{i,\mathrm{th}}&=&f_{i-1}e^{\frac{\sigma_\mathrm{C}}{kT}\left(S_i-S_{i-1}\right)}\simeq
f_{i-1}e^{\frac{2\sigma_\mathrm{C}c_\mathrm{s}v_\mathrm{C}^{2/3}}{3i^{1/3}kT}}=
\nonumber \\ &=& n_\mathrm{C,eq}\lambda c_\mathrm{s}
  \left[(i-1)v_\mathrm{C}\right]^{2/3} \sqrt{\frac{kT}{2\pi
      m_\mathrm{C}}}e^{\frac{2\sigma_\mathrm{C}c_\mathrm{s}v_\mathrm{C}^{2/3}}{3i^{1/3}kT}}
\label{eq:gith}
\end{eqnarray}
where the subscript $_{\mathrm{th}}$ indicates that the above
detachment rate is the rate of spontaneous detachments from a cluster
at temperature $T$.  In the environment of a supernova remnant,
additional processes can enhance the loss of monomers from clusters,
and therefore increase the total ejection rate.  In this work we
consider the chemical reactions between helium cations and carbon
clusters (ion-molecule) and the reaction between a carbon cluster and
an oxygen atom to give a carbon monoxide molecule and a smaller
cluster.  Both reactions have been studied only with very small carbon
clusters $\mathrm{C}_i$ with $i<10$.

The reaction of carbon clusters with oxygen atoms was studied by Woon
\& Herbst (1996) and Terzieva \& Herbst (1998) for carbon clusters up
to $9$ atoms (see also Cherchneff \& Dwek 2010).  They found that the
reaction rate with $i$-sized clusters per unit time and volume
$k_{(\protect\ref{reac:O}),i}$ is given by:
\begin{equation}
k_{(\protect\ref{reac:O}),i}=A_{(\protect\ref{reac:O}),i}\left(\frac{T}{300}\right)^{b_{(\protect\ref{reac:O}),i}}
e^{-\frac{T_{(\protect\ref{reac:O}),i}}{T}}n_{\mathrm{C}_i}n_\mathrm{O}
\label{eq:rateO}
\end{equation}
where the subscript of the coefficients of the modified Arrhenius
equation refer to Eq.~\ref{reac:O}.  Terzieva \& Herbst (1998) find
that the coefficients $A_{(\protect\ref{reac:O}),i}$,
$b_{(\protect\ref{reac:O}),i}$, and $T_{(\protect\ref{reac:O}),i}$
depend on the parity of the cluster size $i$ but do not change
monotonically with it.  This is understood as a consequence of the
fact that the most stable allotrope of $\mathrm{C}_i$ with $i\le9$ is
a linear chain, with dangling bonds at the two extremities
irrespective of the size.  For $10\le{i}\le26$ carbon clusters
organize in rings, while for bigger clusters they form closed
three-dimensional structures (Mauney et al.\ 2015).  In vacuum, big
carbon clusters form fullerenes.  It is believed that astrophysical
carbonaceous grains are mainly in the form of amorphous carbon and/or
graphite onions or flakes.  Either case, it is reasonable to assume
that the number of dangling bonds that can react with free oxygen
atoms grows with the cluster surface.  Under our assumption of
spherical grains, we have therefore that the loss of carbon monomers
from an $i$-sized cluster due to oxygen weathering
is\footnote{Adopting the approximation below or the computed values
  for the small carbon clusters with up to 10 atoms (Terzieva \&
  Herbst 1998) does not result in any significant difference in the
  dust formation yields or carbon monoxide abundance. We show below
  the results from the approximation of Eq~\ref{eq:oxi}.}:
\begin{equation}
g_{i,\mathrm{O}}=10^{-11}e^{-\frac{1130}{T}}n_\mathrm{O,gas}i^{2/3}\;,
\label{eq:oxi}
\end{equation}
where $n_\mathrm{O,gas}$ is the density of gas-phase oxygen atoms.

An analogous discussion holds for the ion-molecule
reaction~\ref{reac:He} whose Arrhenius coefficients for $C_i$ clusters
with $i\le10$ are presented in the UMIST Database for Astrochemistry
2012 (McElroy et al.\ 2013).  In this case the coefficients have no
dependence on $i$, not even with parity.  Again, we assume that the
reaction rate is proportional to the number of dangling bonds which,
in turn, scales with the grain surface. We have therefore that the
loss of carbon monomers from an $i$-sized cluster due to HeII
weathering is:
\begin{equation}
g_{i,\mathrm{He}}=1.6\times10^{-9}n_\mathrm{HeII}i^{2/3}
\end{equation}
where $n_\mathrm{HeII}$ is the density of singly ionized helium atoms.

The total ejection rate that we use in Eq.~\ref{eq:J} is therefore the
sum of the spontaneous ejection rate plus the chemical weathering due
to oxygen and ionized helium:
\begin{equation}
g_i=g_{i,\mathrm{th}}+g_{i,\mathrm{O}},+g_{i,\mathrm{He}}
\label{eq:gi}
\end{equation}
The consequence of this equation is that the ejection rate is
increased, making nucleation of new grains more difficult.  The
spontaneous ejection rate allows for nucleation as soon as the
saturation is larger than one, and nucleation becomes efficient for
saturation larger than a few.  As we will see in the following, in the
environment of a SNR chemical weathering of the forming grains is so
efficient that nucleation only takes place for saturations of many
orders of magnitude larger than unity.

No matter how hampered by chemical weathering, the nucleation of small
grains still takes place, albeit at a much reduced rate.  Once stable
grains are formed (where stable means bigger than the critical size at
which $f_i=g_i$, their size evolves according to the same accretion
and decretion rate discussed above.  Such critical size is a function
of temperature and saturation and ranges between approximately
$3\,$\AA{} and $30\,$\AA.  The growth of existing grains of size $i$
in monomers per second is regulated by the equation
\begin{equation}
\left.\frac{dn}{dt}\right|_i=f_i-g_i\;.
\label{eq:growth}
\end{equation}
Note that the stability of grains can change as a result of the
cooling of the gas (increasing stability) and of the decrease of
gas-phase C atoms (decreasing stability) and therefore a grain that is
formed as stable at some time can later become unstable and shrink in
size.

\section{Supernova model}

\begin{figure}
\includegraphics[width=\columnwidth]{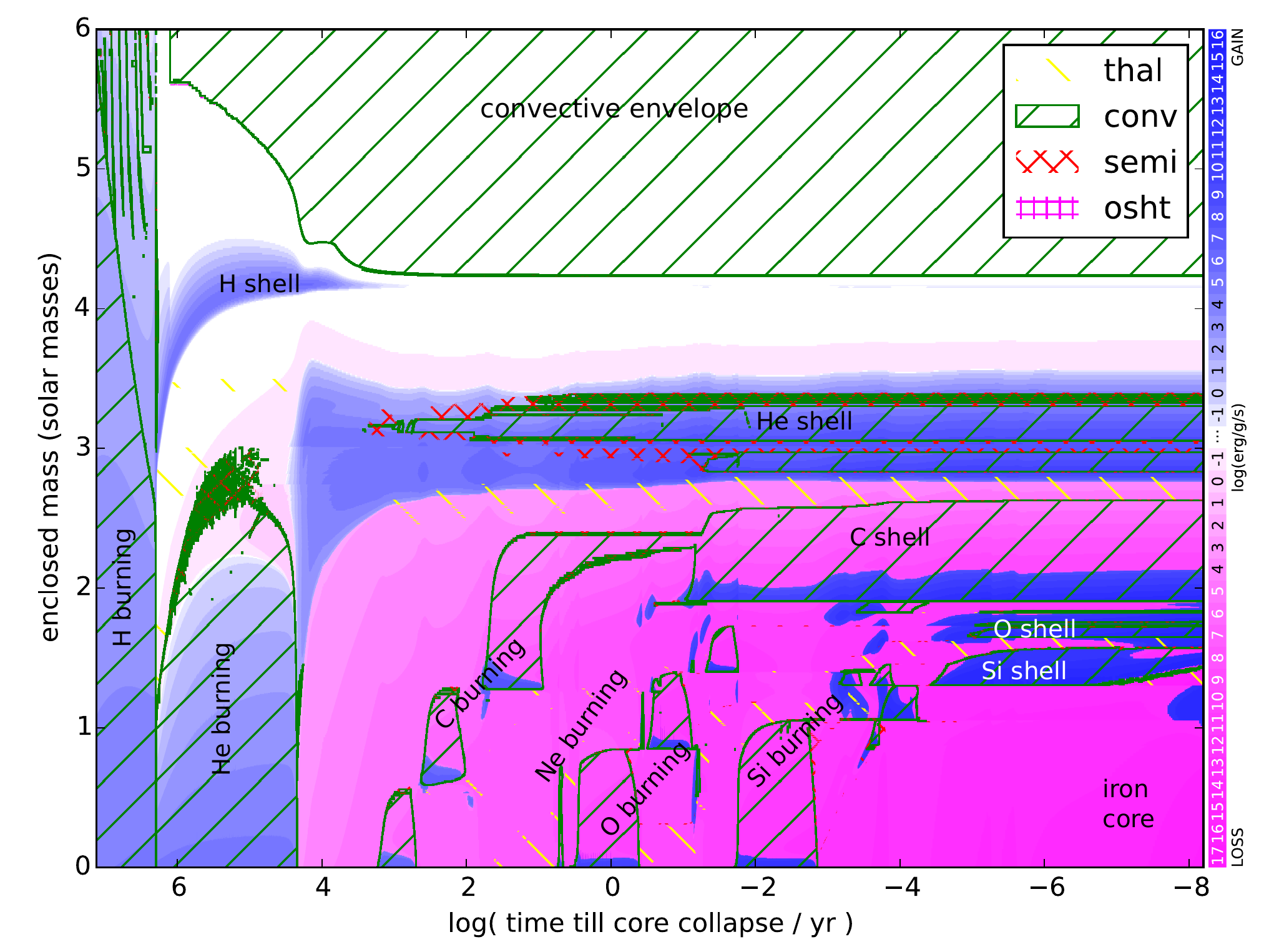}
\caption{Kippenhahn diagram of the pre-supernova evolution of the
  \textbf{inner $6\,\Msun$ of a} $15\,\Msun$ progenitor star.  The
  \textsl{x-axis} shoes the time till core collapse in yr, from $30$
  million yr before collapse to $250\,$ms before core bounce; the
  \textsl{y-axis} is the mass coordinate; \textsl{blue shading}
  indicate regions where nuclear energy generation exceeds neutrino
  losses, \textsl{purple shading} where neutrino losses dominate;
  \textsl{green hatching} (framed) indicates convective regions,
  \textsl{horizontal/vertical purple hatching} indicates convective
  overshooting (not visible), \textsl{red cross hatching} indicates
  semi-convective regions, and \textsl{yellow hatching} thermohaline
  mixing.  \textbf{Major nuclear burning stages are indicated.}
\label{fig:s15_cnv}}
\end{figure}

\begin{figure}
\includegraphics[width=\columnwidth]{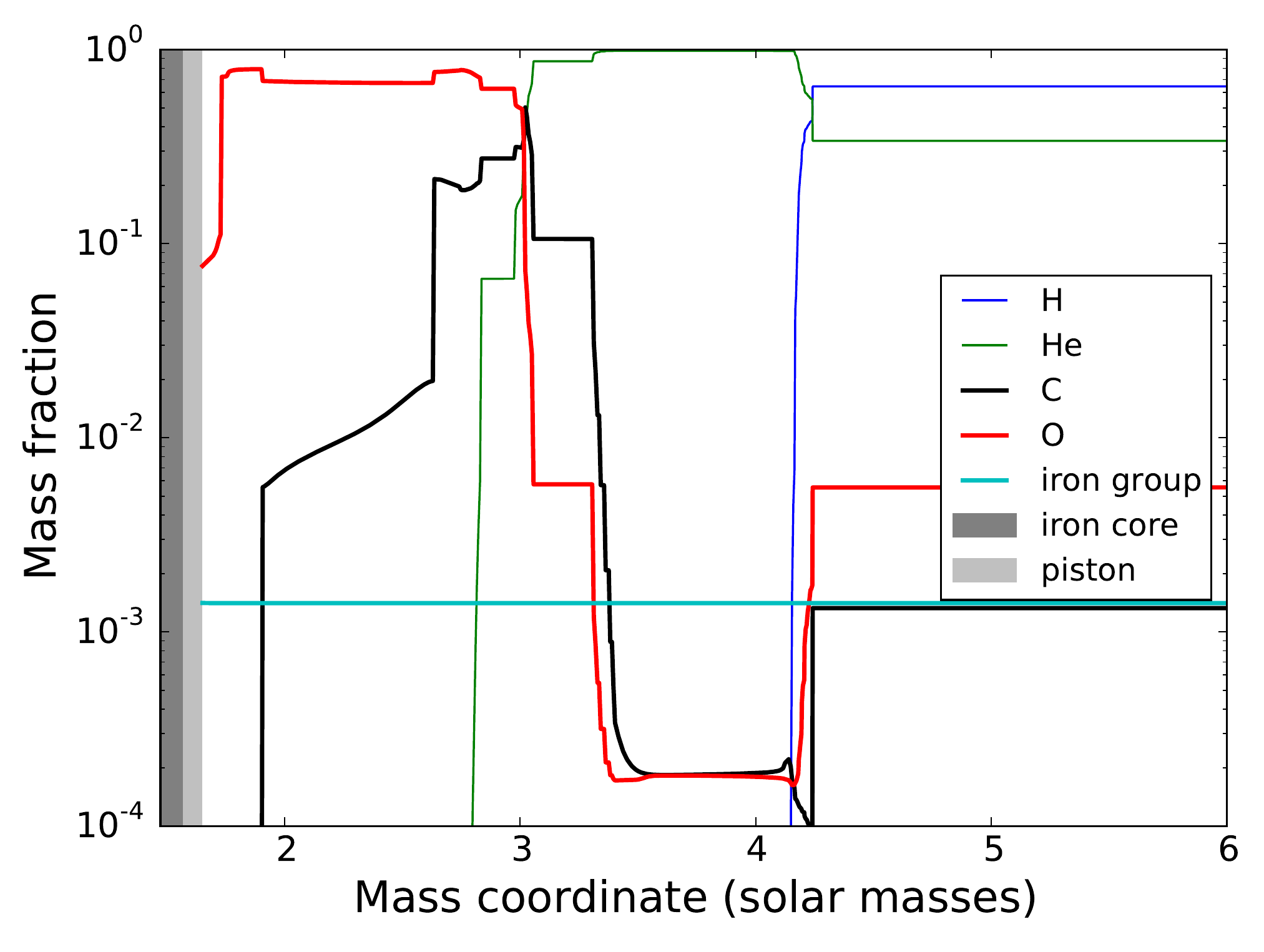}
\caption{Presupernova structure of the $15\,\Msun$ model.
  \textsl{Dark gray shading} indicates the iron core and \textsl{light
    gray shading} the location of the piston (initial mass cut).
  \textbf{The mass coordinate has been truncated on both sides.  The
    envelope is chemically homogeneous above $6\,\Msun$ and the total
    mass of the star at the time of supernova is $11.185\,\Msun$.}
\label{fig:s15_presn}}
\end{figure}

\begin{figure}
\includegraphics[width=\columnwidth]{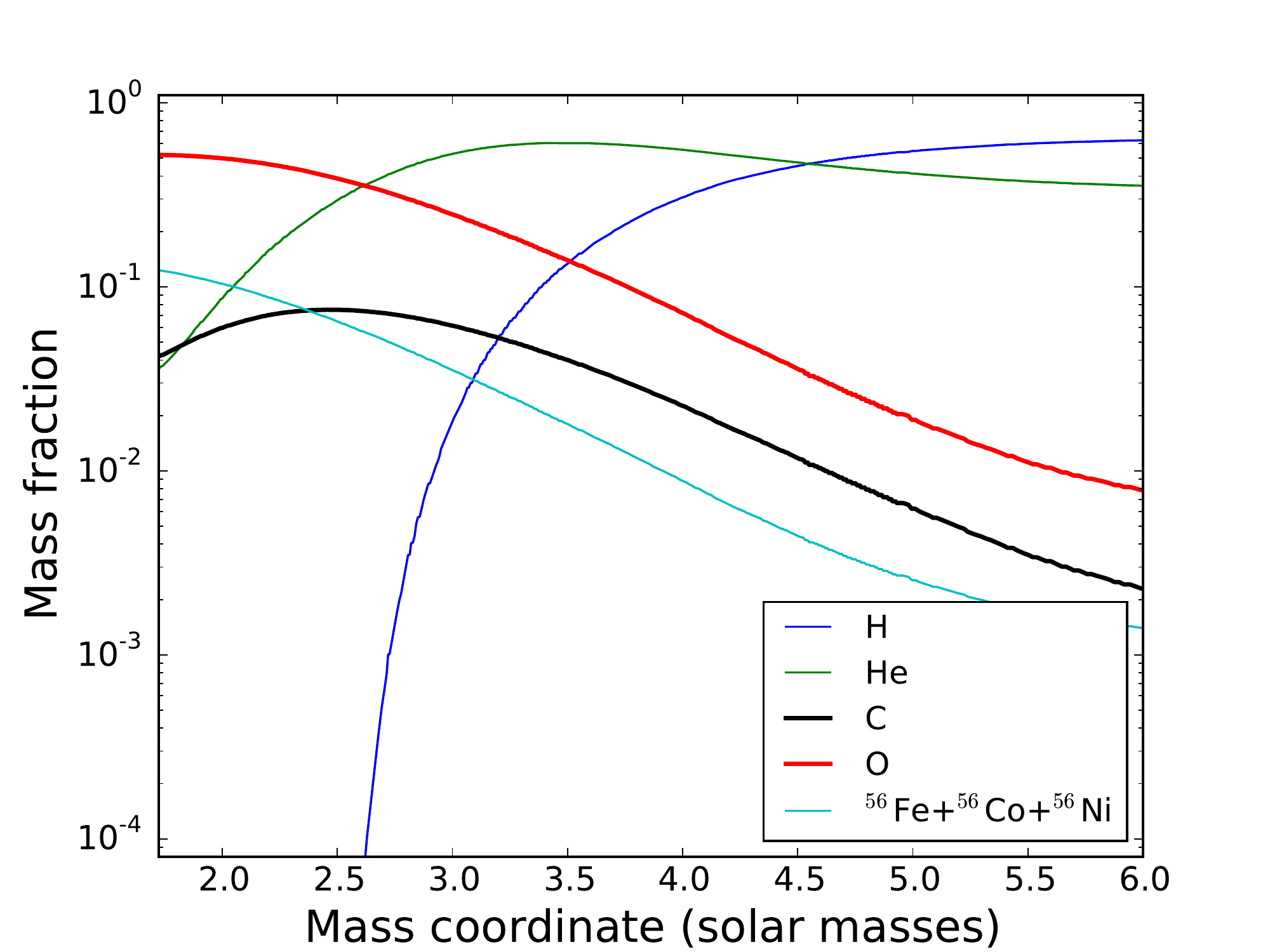}
\caption{{Post-explosion mass fraction of the elements relevant to
  carbonaceous dust formation as a function of the inner enclosed mass
  for the mixed stellar progenitor.}
\label{fig:m_massfrac}}
\end{figure}

We simulated a $15\,\Msun$ solar model with initial solar composition
of Lodders et al.\ (2009) from the zero-age main sequence to pre-SN
phase using the KEPLER (Weaver et al.\ 1978) stellar evolution code.
We use the same prescription of hydrodynamic instabilities
(semiconvection, overshoot, convection mixing length parameter),
nuclear reaction rates, and mass loss rates as in Woosley \& Heger
(2007).  Figure~\ref{fig:s15_cnv} shows the Kippenhahn Diagram of the
pre-supernova evolution, and Figure~\ref{fig:s15_presn} the
composition of the model at pre-supernova stage.  We simulate the
explosion using a hydrodynamic spherically symmetric piston that is
adjusted to yield a total kinetic energy of the ejecta of
$1.2\times10^{51}\,$erg (see Rauscher et al.\ 2002 for a more detailed
description).  The post-SN mixing due to Rayleigh-Taylor instabilities
is approximated by a simple ``boxcar'' model and adjusted to mix
$^{56}$Ni out into the envelope such that typical SN light curves are
reasonably reproduced (Figure~\ref{fig:m_massfrac}, see also Rauscher
et al.\ 2002).  The amount of mixing is consistent with hydrodynamic
simulations (e.g., Joggerst et al.\ 2010).

\begin{figure}
\includegraphics[width=\columnwidth]{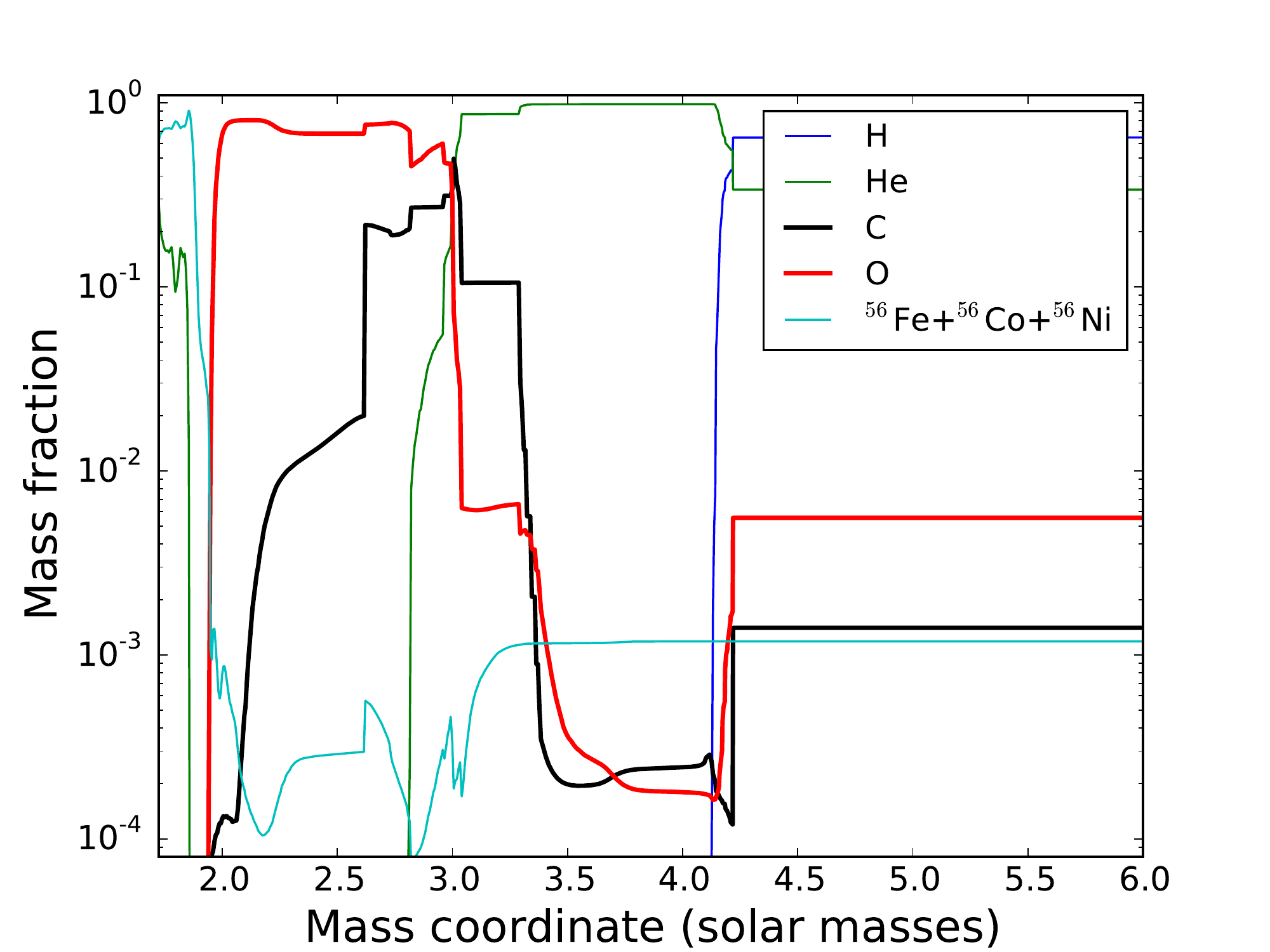}
\caption{{Same as Figure~\ref{fig:m_massfrac} but for the unmixed
    model.}
\label{fig:unm_massfrac}}
\end{figure}

Note that the $^{56}$Ni present in the innermost layers seen in
Figure~\ref{fig:unm_massfrac} is made in the supernova explosion by
shock heating.  The $^4$He seen in this region is due to
photo-disintegration by the shock and incomplete recombination to
$^{56}$Ni.  For reference, this is the same $15\,\Msun$ model also
included in Patnaude et al.\ (2015).

\begin{table}
\caption{Properties of representative zones.\label{zones}}
\begin{tabular}{lrrrr}
\hline\hline
\multicolumn{1}{l}{Zone}&\multicolumn{1}{r}{Zone}
&\multicolumn{1}{r}{Mass Coordinate} &
\multicolumn{1}{r}{$n_\mathrm{C}/n_\mathrm{O}$}& \multicolumn{1}{r}{$n_\mathrm{C}/n_\mathrm{O}$} \\
\multicolumn{1}{l}{Name}& \multicolumn{1}{r}{\#}
&\multicolumn{1}{r}{(Solar Masses)}  & \multicolumn{1}{r}{mixed}& \multicolumn{1}{r}{unmixed} \\
\hline
\zone{69}   & 69  & 2.11            & 0.18 &  0.0016\\
\zone{209}  & 209 & 3.0            &  0.22 & 1.44 \\
\zone{556}  & 556 & 5.0            & 0.44 & 0.32 \\
\hline\hline
\end{tabular}\\
$^a$ logarithm base 10 of number ratio
\end{table}

\begin{figure}
\includegraphics[width=\columnwidth]{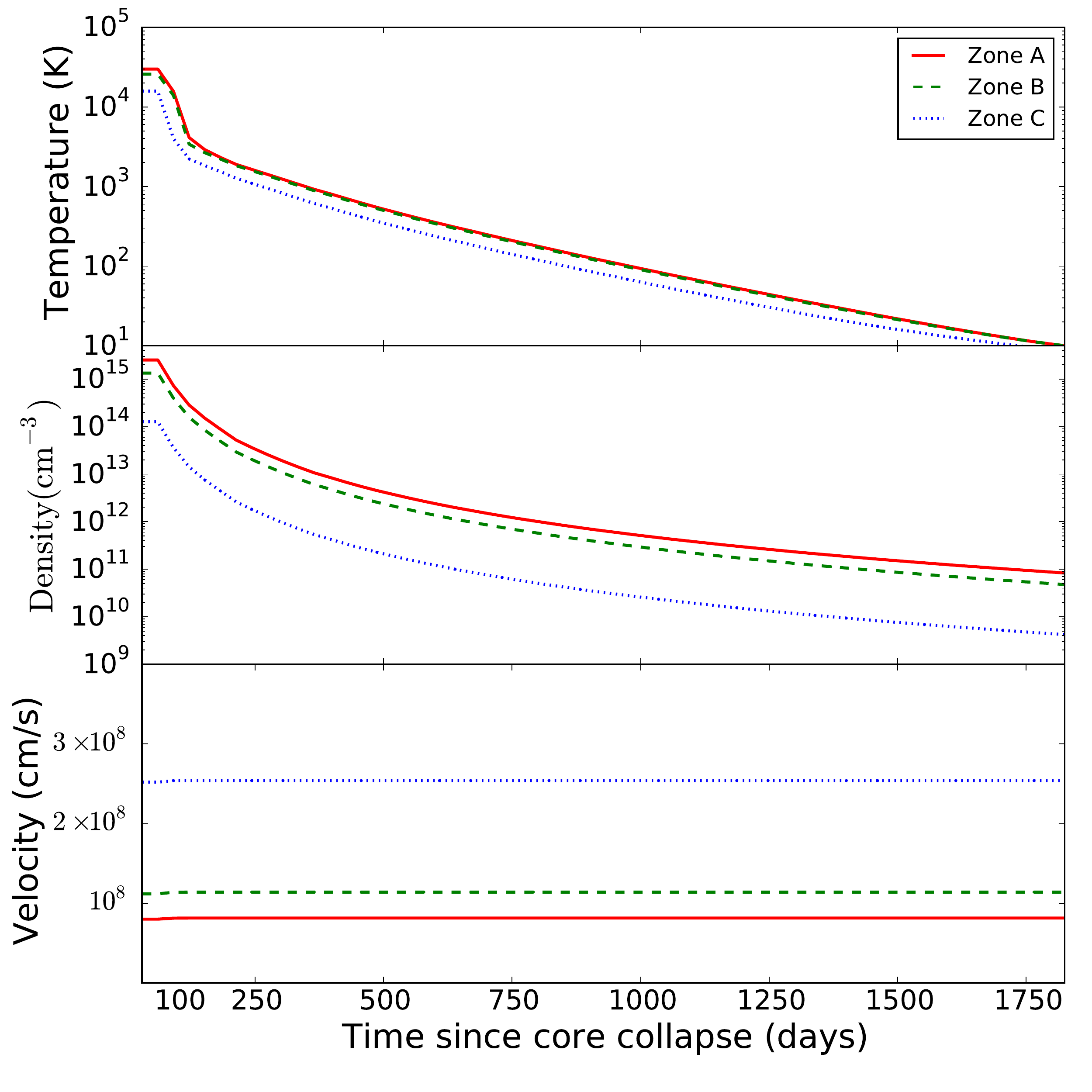}
\caption{ Temporal evolution of the dynamical properties of the three
  zones for which the dust formation and chemical evolution is
  discussed in more detail in the text.
  See Table~\ref{zones} for more details on the three zones.
  \label{fig:dynevol}}
\end{figure}

\section{Results}

\begin{figure}
\includegraphics[width=\columnwidth]{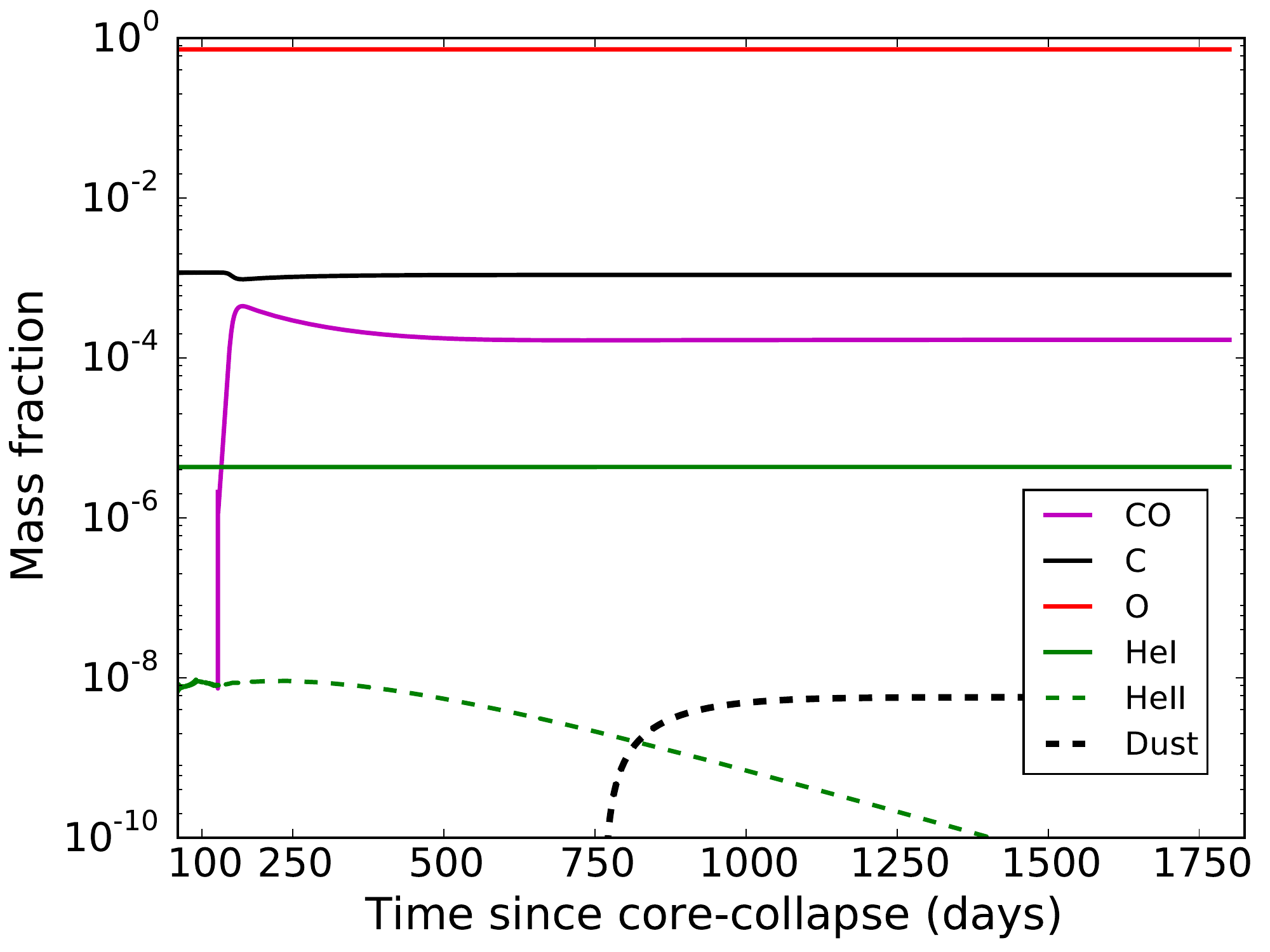}
\caption{Mass fraction of relevant species (atoms, ions, molecules, or
  dust) as a function of time in the Simulation \protect\Zone{69}
  (Table~\ref{zones}).
\label{fig:69_mol}}
\end{figure}

In this section we present the results of our calculations focusing on
the differences between results from the carbon nucleation, growth and
weathering presented here with respect to more traditional
calculations based on classical nucleation theory and alternative
approaches that aim at taking into account the chemistry involved in
the formation of the small dust seeds.  Let us first look in detail at
the temporal evolution of the properties of a few representative
single shells.  We chose three representative shells zones
(Table~\ref{zones}) in the unmixed stellar progenitor model
(Figure~\ref{fig:unm_massfrac}).  \Zone{69} is located at the
$2\,\Msun$ mass coordinate, and is characterized by a marked oxygen
overabundance with respect to carbon (approximately three orders of
magnitude).  \Zone{209} is located at mass coordinate $3\,\Msun$ and
has approximately equal abundances of C and O.  It is also the most
carbon-rich zone in the whole ejecta.  Finally, \Zone{556} is located
at mass coordinate $5\,\Msun$ well outside the region were explosive
nucleosynthesis was active. It is characterized by an overabundance of
O by a factor $\sim3$ with respect to C. The thickness of each zone is
small and within each of them the densities and thermodynamic
properties are constant at any given time. The dynamical evolution of
the three zones (temperature, density, and expansion velocity) is
shown in Figure~\ref{fig:dynevol}.

\begin{figure}
\includegraphics[width=\columnwidth]{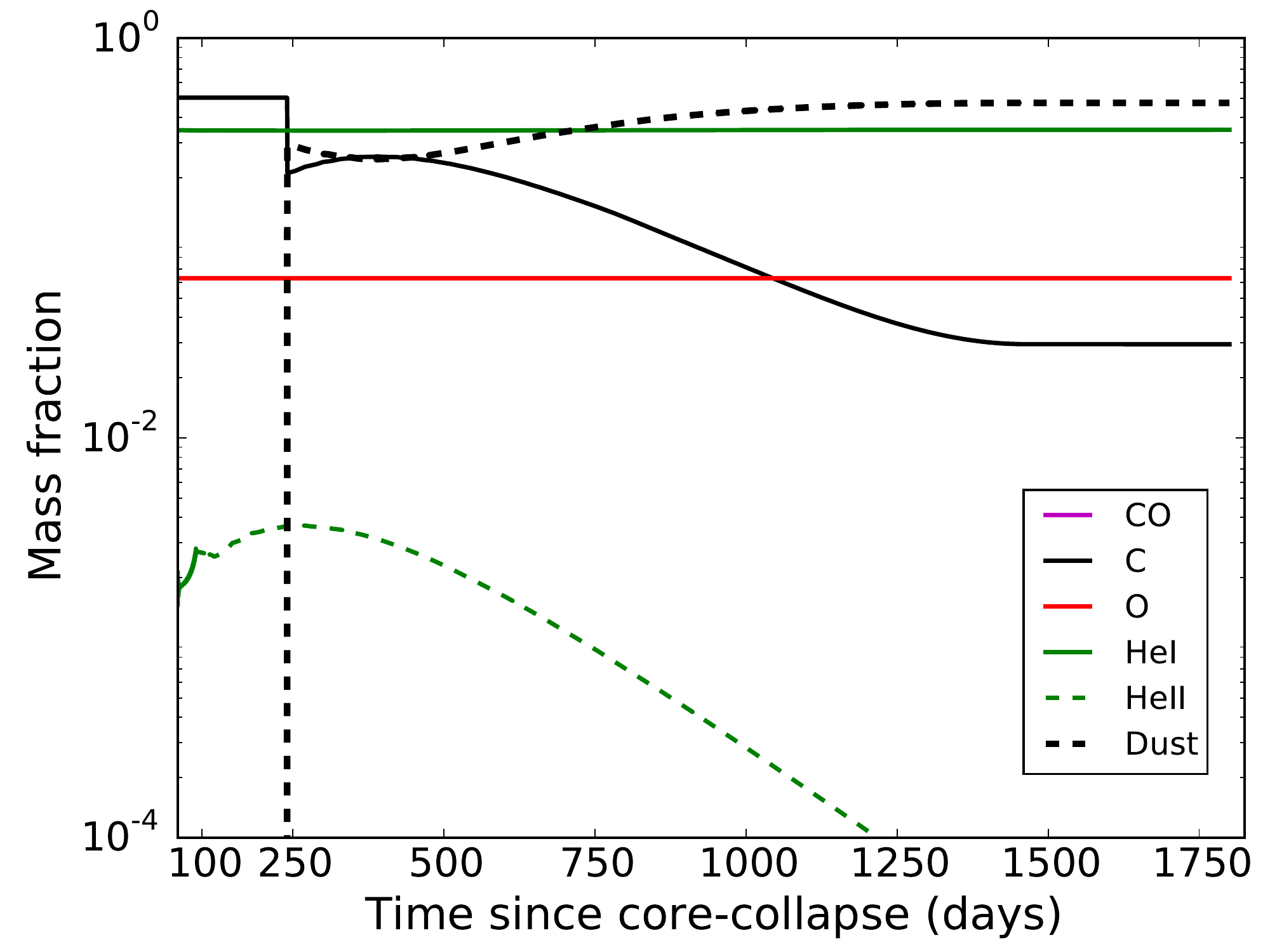}
\caption{Same as Figure~\ref{fig:69_mol} but for \protect\Zone{209}
    (Table \ref{zones}).
\label{fig:209_mol}}
\end{figure}

\begin{figure}
\includegraphics[width=\columnwidth]{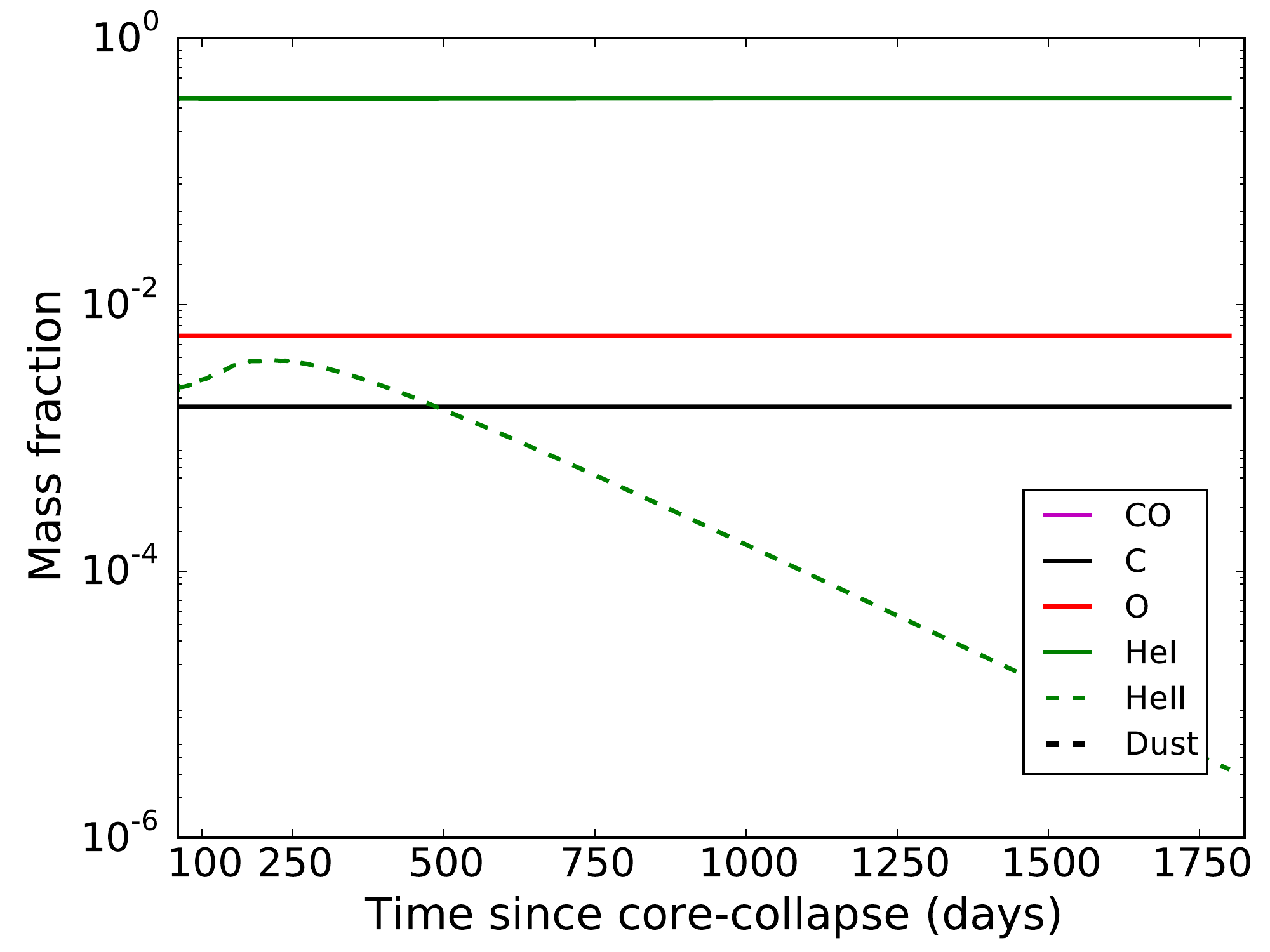}
\caption{Same as Figure~\ref{fig:69_mol} but for \protect\Zone{556}
  (Table~\ref{zones}).
\label{fig:556_mol}}
\end{figure}

Figures~\ref{fig:69_mol}, \ref{fig:209_mol}, and~\ref{fig:556_mol}
show the temporal evolution of the ejecta composition in the three
selected zones for the unmixed progenitor model.  Despite some
differences, all the plots show some important common features.
First, the formation of the CO molecule is never efficient enough to
completely absorb all the carbon, even in \Zone{69}, where oxygen is
largely overabundant with respect to carbon (see analogous results by
Clayton et al.\ 2001; Deneault et al.\ 2006, and Clayton 2013).
Second, a sizable fraction of He is ionized to the $\mathrm{He}^+$
cation, and its weathering effect can delay dust formation.  This is
particularly clear in \Zone{209}, where dust initially forms quickly
at $t\sim7$ months after the explosion.  The $\mathrm{He}^+$
concentration, however, peaks a few months later causing a temporary
decrease in the dust fraction.  Eventually, the ionized He fraction
drops again and dust nucleation and growth resume.  This result not
only shows the important effects of weathering in the supernova
environment, it also shows that the customary assumption that once a
grain has grown beyond the critical size it only grows further is
inaccurate.  Weathering can become dominant and shrink or completely
destroy the already formed grains.  Oxidation is also important in
limiting the dust formation and growth. Its effect, however, is less
clear in the figures because the O abundance is constant with time.
The rate of the He-C$_i$ ion-molecule reaction is also approximately
two orders of magnitude faster than the oxidation reaction, making the
cation the most damaging weathering agent in all regions except for
those in which oxygen is particularly abundant.

Figures~\ref{fig:69_mol}, \ref{fig:209_mol}, and~\ref{fig:556_mol}
also show that the dust condensation time and its efficiency vary
significantly across the remnant.  Condensation is significantly
delayed in the inner mass coordinate, starting only $\sim$2 years
after core-collapse.  Mostly due to atomic carbon depletion into CO,
the dust condensation efficiency is very small in \Zone{69}.  In the
carbon-rich area of \Zone{209}, instead, dust formation sets in early,
but is subsequently reversed into dust destruction by the increasing
ionization of helium.  Only a few years after core collapse dust
formation reaches its final efficiency, almost $100\,\%$.  Finally, no
dust is formed in the outer layers of the star (\Zone{556}).

\begin{figure}
\includegraphics[width=\columnwidth]{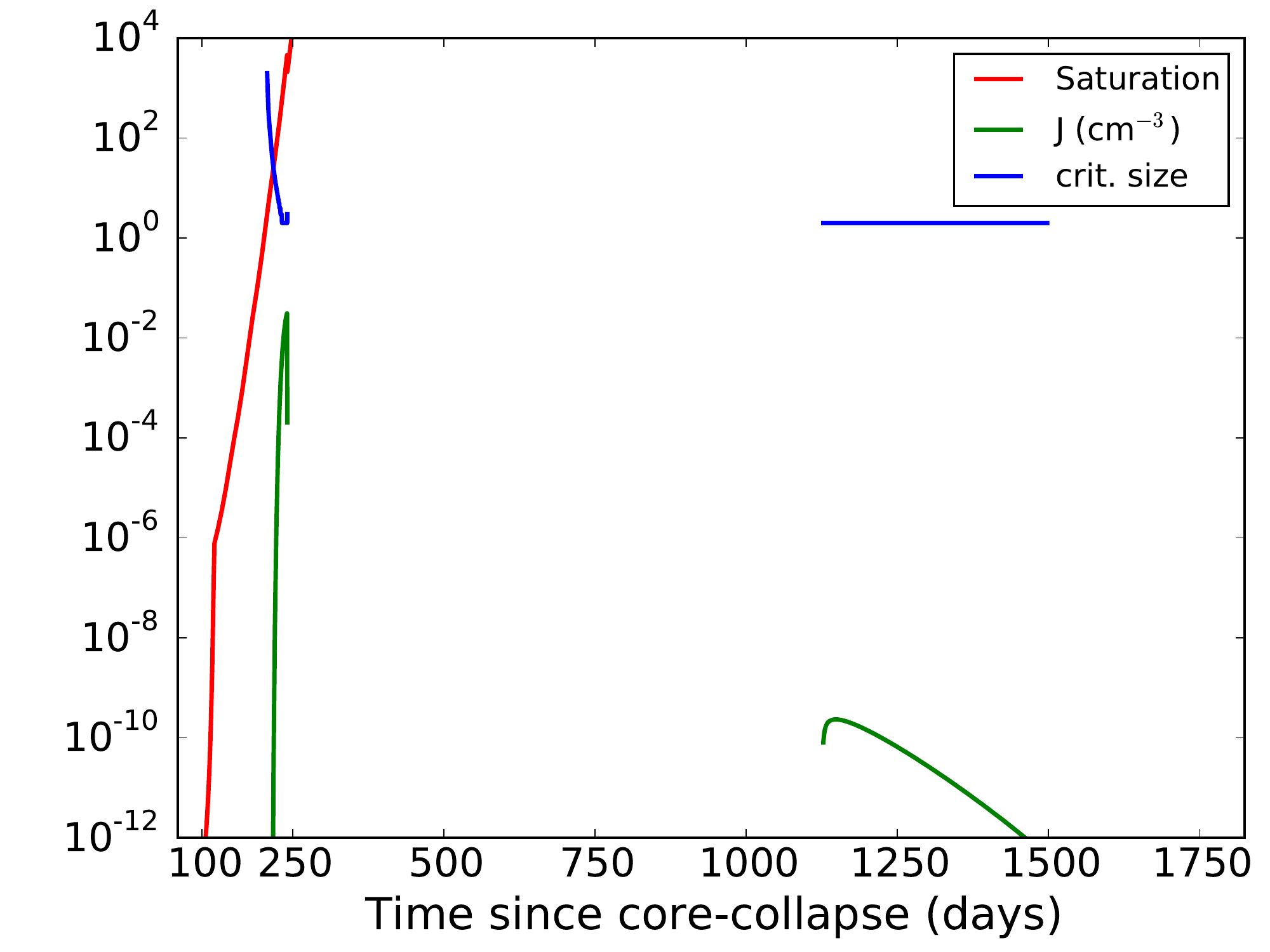}
\caption{Evolution of the saturation ($S$), nucleation rate ($J$) and
  critical size of \protect\Zone{209} (Table~\ref{zones}).
\label{fig:J_209}}
\end{figure}

A fundamental role in altering the nucleation pattern of carbonaceous
dust is also played by the difficulty in forming the dimer, as
discussed above.  Figure~\ref{fig:J_209} shows the saturation,
nucleation rate, and critical size of the forming carbon grains as a
function of time for \Zone{209}.  We first note that saturation does
not play the same role that it does in the thermodynamic theory of
homogeneous nucleation.  In that case, the solid phase is expected to
start nucleating as soon as saturation grows above a few.  In our
calculation, saturation can grow to extremely high values, as large as
a double precision real number can handle.  Yet, nucleation proceeds
slowly, and can even be turned off while the saturation is increasing.
What we see in the figure is that initially, as saturation becomes
larger than one, a spike of nucleation is observed (at
$t\sim7$~months).  During this phase the spontaneous ejection of
monomers from unstable carbon clusters dominates over the oxidation
and charge-exchange rates.  As usual, in this phase the growth of the
saturation produces an increased nucleation rate and smaller critical
size.  Differently from classical calculations, however, our
nucleation rate is limited by the dimer formation rate
(Eq.~\ref{eq:C2}).

\begin{figure}
\includegraphics[width=\columnwidth]{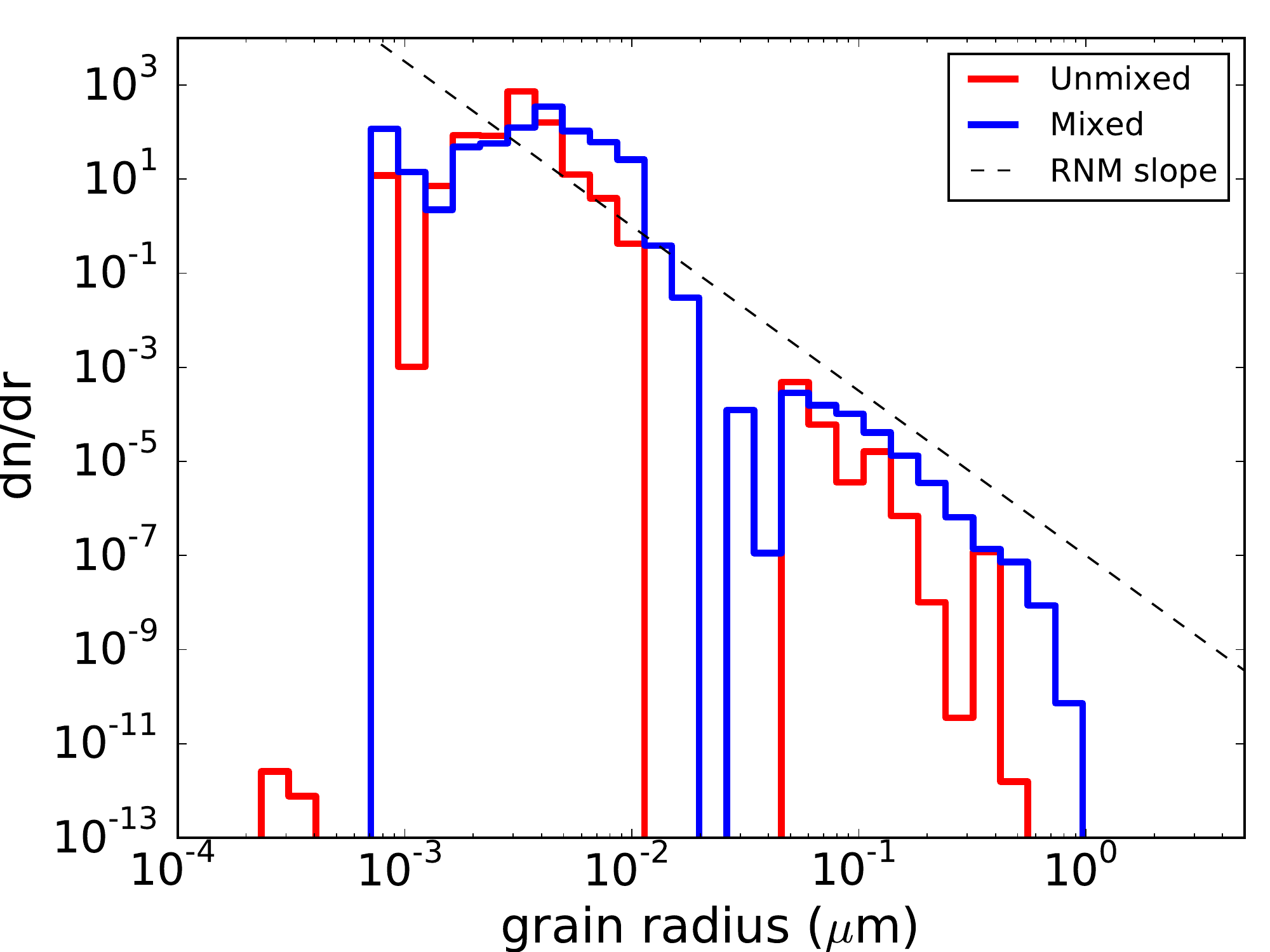}
\caption{Size distribution of the carbon grains for the Mixed and
  Unmixed models integrated over the whole ejecta at 60 months after core-collapse.
  \label{fig:hist}}
\end{figure}

As time progresses, the concentration of HeII cations increases and
the charge-exchange weathering process becomes dominant, shutting off
entirely the nucleation of new grains and shrinking the already
existing ones (see the decreasing dust mass fraction between $\sim7$
and $\sim13$~months in Figure~\ref{fig:209_mol}).  In this period, the
saturation becomes virtually infinite.  Yet, nucleation is entirely
shut off because the weathering rate is larger than the accretion rate
of carbon atoms on any size grain.  At later times ($t>35$~months) the
charge-exchange weathering still dominates the ejection rate, but the
accretion of carbon atoms now dominates the rates.  Nucleation can
therefore proceed, with a formal critical size of two atoms.  Again,
the nucleation rate is constrained by the rate of the formation of the
dimer.  The plots for the other two zones are analogous, but with only
a subset of the phenomenology of what seen in \Zone{209}.  The inner
\Zone{69} does not have enough carbon to trigger the initial burst of
dust formation, and nucleation is postponed until $t\sim2$~years, when
the ion-molecule weathering rate becomes smaller than the accretion
rate.  The nucleation takes place very slowly, being limited by the
rate of dimer formation.  For the outer \Zone{556}, instead,
nucleation never takes place, because the carbon abundance is so low
compared to the He abundance that charge-exchange weathering is always
dominating over the growth of carbon clusters.

Let us now look at nucleation and growth of carbonaceous dust in the
whole ejecta.  Figure~\ref{fig:hist} shows the comparison of the final
grain size distributions at 5 years past the core collapse for the
mixed and unmixed models.  Both distributions have been normalized to
unit integral to facilitate the comparison and a dashed line has been
added to show the classical MRN interstellar size distribution (Mathis
et al.\ 1977).  In both cases we obtain a broad size distribution that
is somewhat steeper than the MRN distribution. The Mixed progenitor
model gives a bigger upper limit for the grain size. Since the gain
population still needs to cross the reverse shock, where the smallest
grains are likely to be destroyed by sputtering (Nozawa et al.\ 2006;
Bianchi \& Schneider 2007), the small differences seen here could be
much more substantial when the dust is recycled into the interstellar
medium.

\begin{figure}
\includegraphics[width=\columnwidth]{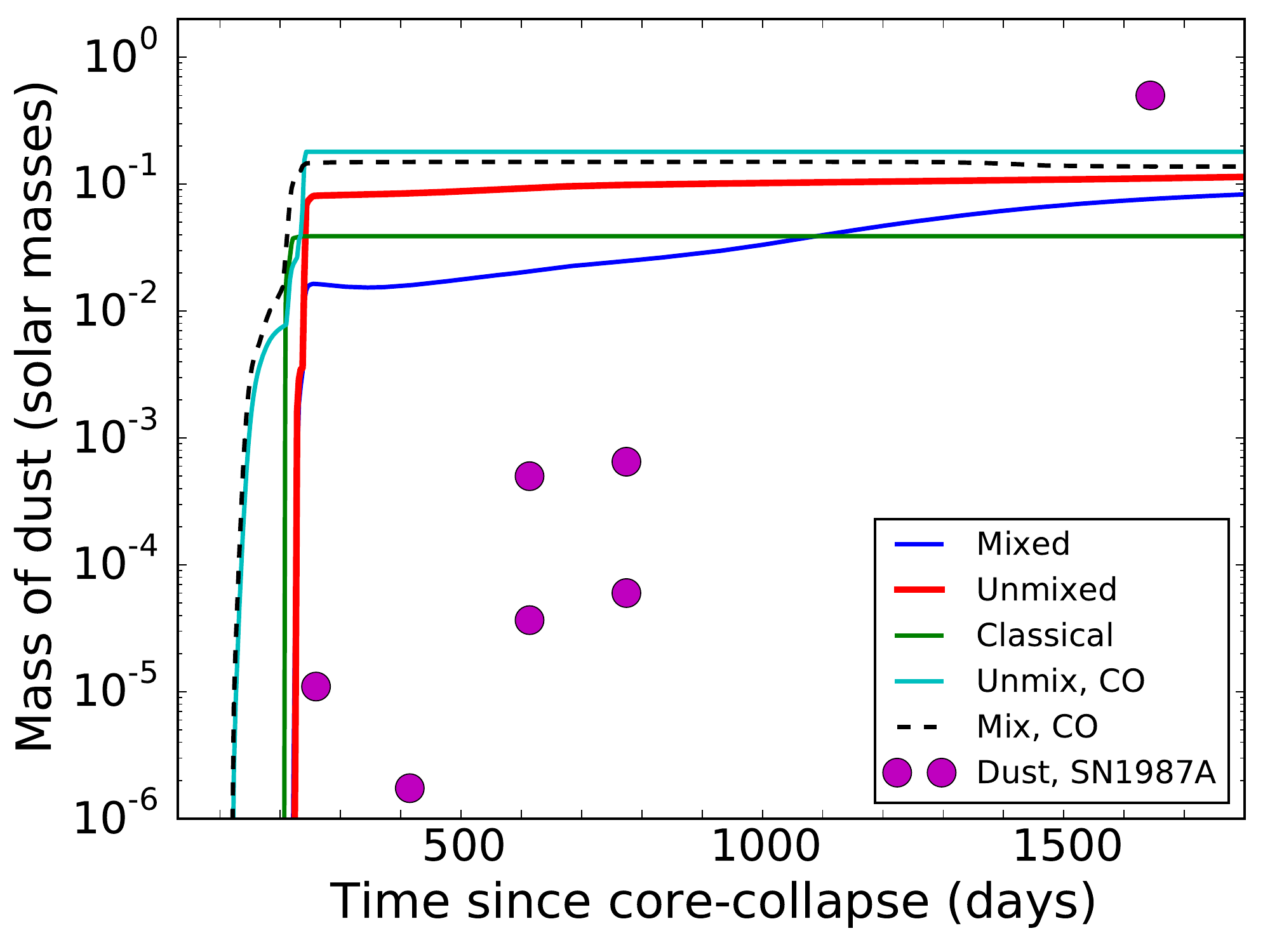}
\caption{Evolution of the total mass of carbonaceous grains since the
    core-collapse. Various dust formation models are used and are
    shown in different colors, as discussed in the text.
\label{fig:dust_mass}}
\end{figure}

In the following, we consider a suite of codes with different
assumptions to compare and understand the effect of adding new
processes in the dust formation process.  The first code, which is
labeled as ``Classical'' in the figures, adopts the standard
assumptions of classical calculations.  In this version of the code,
nucleation takes place only in regions where the number density of
carbon atoms is larger than that of oxygen atoms. The nucleation rate
is computed according to the thermodynamic theory as (e.g., Fallest et
al.\ 2011 and references therein):
\begin{equation}
J=\left(\frac{c_\mathrm{s}^3v_C^2\sigma}{18\pi^2m_\mathrm{C}}\right)^{1/2}\,n_\mathrm{C}^2\,
e^{-\frac{4c_\mathrm{s}^3v_\mathrm{C}^2\sigma^3}{27(kT)^3(\ln{S})^2}}
\end{equation}
where we adopt the value $\sigma=1500\,$dyne/cm for the amorphous
carbon surface energy.  Within this theory, the critical grain size
can be found analytically as
\begin{equation}
n^*=\frac{8c_\mathrm{s}^3v_\mathrm{C}^2\sigma^3}{27(kT\ln{S})^3}
\end{equation}
This dust model was applied only to the unmixed progenitor, since the
mixed one does not have any zone in which $n_\mathrm{C}>n_\mathrm{O}$.

\begin{figure}
\includegraphics[width=\columnwidth]{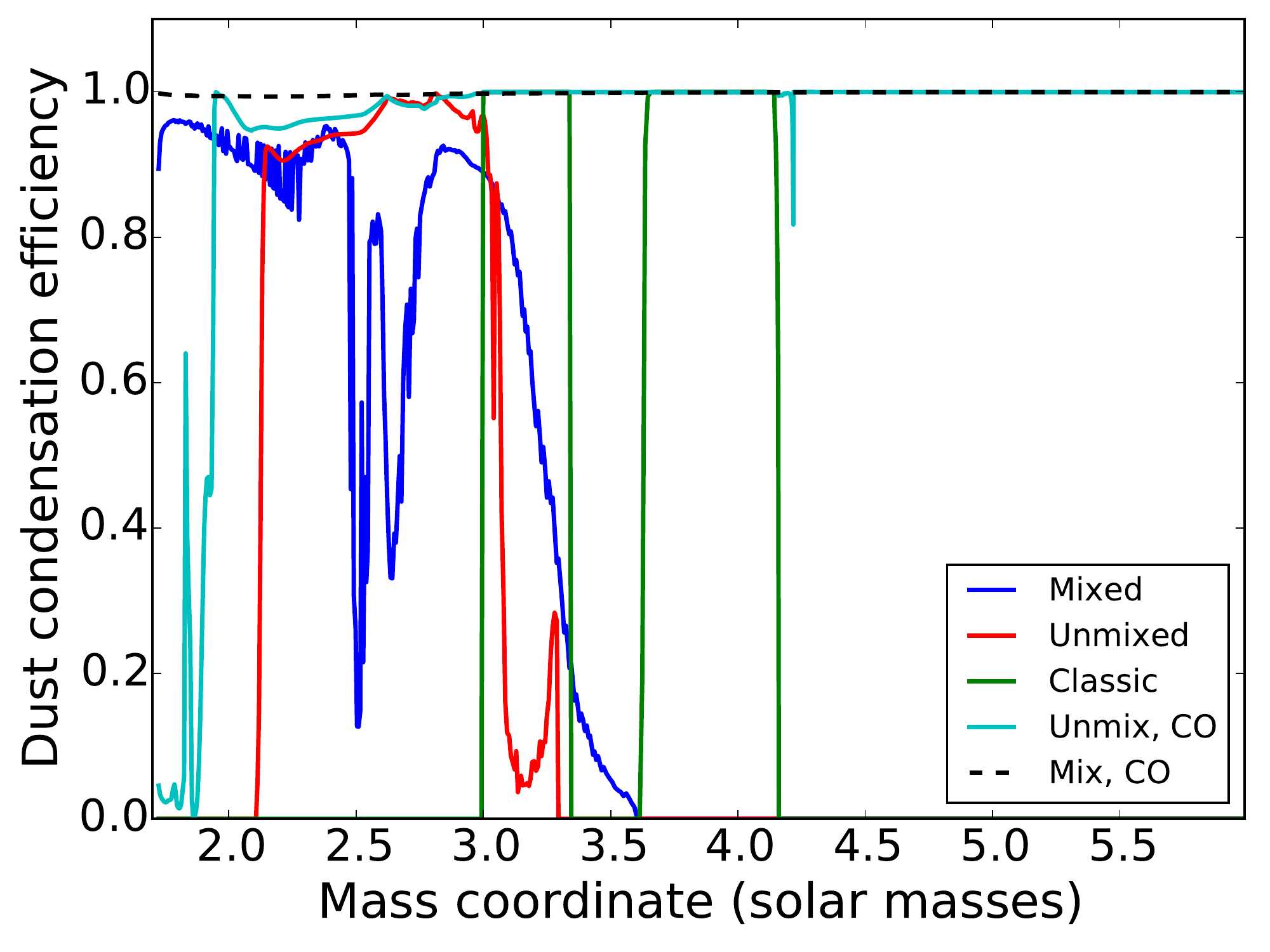}
\caption{Efficiency of the carbonaceous dust condensation at Year 5
  (the last time of our simulations) as a function of the mass
  coordinate.  The efficiency is defined as the mass in dust grains
  over the total carbon mass.
\label{fig:effi}}
\end{figure}

\begin{figure}
\includegraphics[width=\columnwidth]{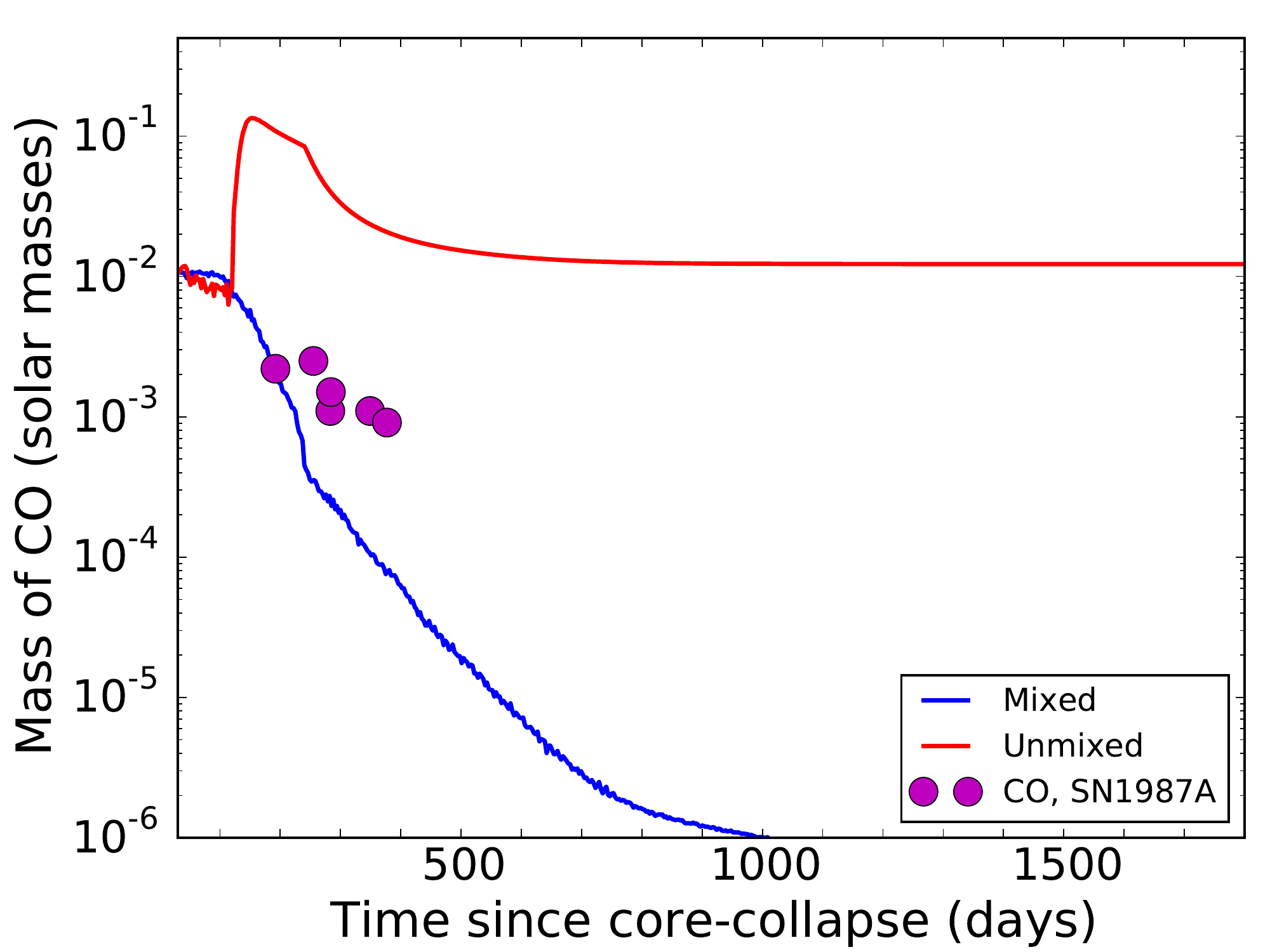}
\caption{Evolution of the total mass of CO as a function of time since the
    core-collapse. The various models are
    shown in different colors, as discussed in the text.
    \label{fig:CO_mass}}
\end{figure}

 The second model we adopt is a slight modification of the
``Classical'' one, in which the formation and destruction of the CO is
followed as described in Sect.~\ref{sec:CO}.  This allows for the
formation of carbonaceous dust even in regions with
$n_\mathrm{C}\le{}n_\mathrm{O}$, and therefore the model is run on
both our mixed and unmixed progenitor models.  The results are labeled
in the figures as ``Unmix, CO'' and ``Mix, CO''.  Finally, our full
code described in Sect.~2 was applied to both the mixed and unmixed
progenitor models.  The results are labeled in the figures as just
``Mixed'' and ``Unmixed''.

Figure~\ref{fig:dust_mass} shows the time evolution of the total mass
condensed in carbonaceous grains.  As expected, we see that all models
based on classical nucleation have a sharp rise in the dust mass, due
to a catastrophic dust formation event at about $7$ months after core
collapse.  The dust mass is subsequently constant.  This is in strong
contrast with the continuum dust condensation detected in a sample of
CCSN (Gall et al.\ 2014) and shown with the magenta symbols that
represent dust detections in the remnant of SN1987A (Wooden et al.\
1993; Ercolano et al.\ 2007; Matsuura et al.\ 2013; Indebetouw et al.\
2014).  The ``Mixed'' and ``Unmixed'' models, instead, show a more
continuous formation of dust, starting approximately at $7$ months (as
for the classical models).  Yet, even the more progressive dust
formation of the Mixed and Unmixed models including weathering is
still too fast to fully reproduce the observations.  Dust formation in
our simulations is continuously active for at least $5$ years after
core collapse, after which our calculations were halted.  Extending
the calculations for longer times is possible but not simple, since
the thermal balance of the remnant becomes much more complex.
Condensation is not expected to proceed much further, however, since
in all the dust formation zones most of the available carbon has
already been used by Year $5$.  This is shown in
Figure~\ref{fig:effi}, where the efficiency is plotted for the various
models as a function of the mass coordinate.  Efficiency here is
defined as the ratio of the mass of carbon in grains over the total
mass of carbon.  Where dust formation takes place, it is almost
inevitably highly efficient ($\sim1$), converting the vast majority of
carbon into solid grains.  This is not a trivial result.  Consider,
for example, model ``Mix, CO''.  In all the zones oxygen is
overabundant with respect to carbon. One might think, therefore, that
most of the carbon should remain locked in CO molecules instead of
condensing in carbonaceous grains.  Instead, we see that while the CO
molecule does form, it is continuously dissociated by Compton
electrons.  Grains instead, once they have managed to grow beyond a
certain size, are much more resilient due to the fact that only the
surface can be weathered by oxidation and ion-molecule reactions.  In
addition, Compton electrons are not damaging to grains because of the
huge mass difference.  Classical models with CO
formation/dissociation, therefore, have virtually $100\,\%$
condensation efficiency of all carbon into grains across the entire
star. The classical model without CO dissociation, instead, has
$100\,\%$ efficiency but only in those zones where
$n_\mathrm{C}>n_\mathrm{O}$.

\begin{figure}
\includegraphics[width=\columnwidth]{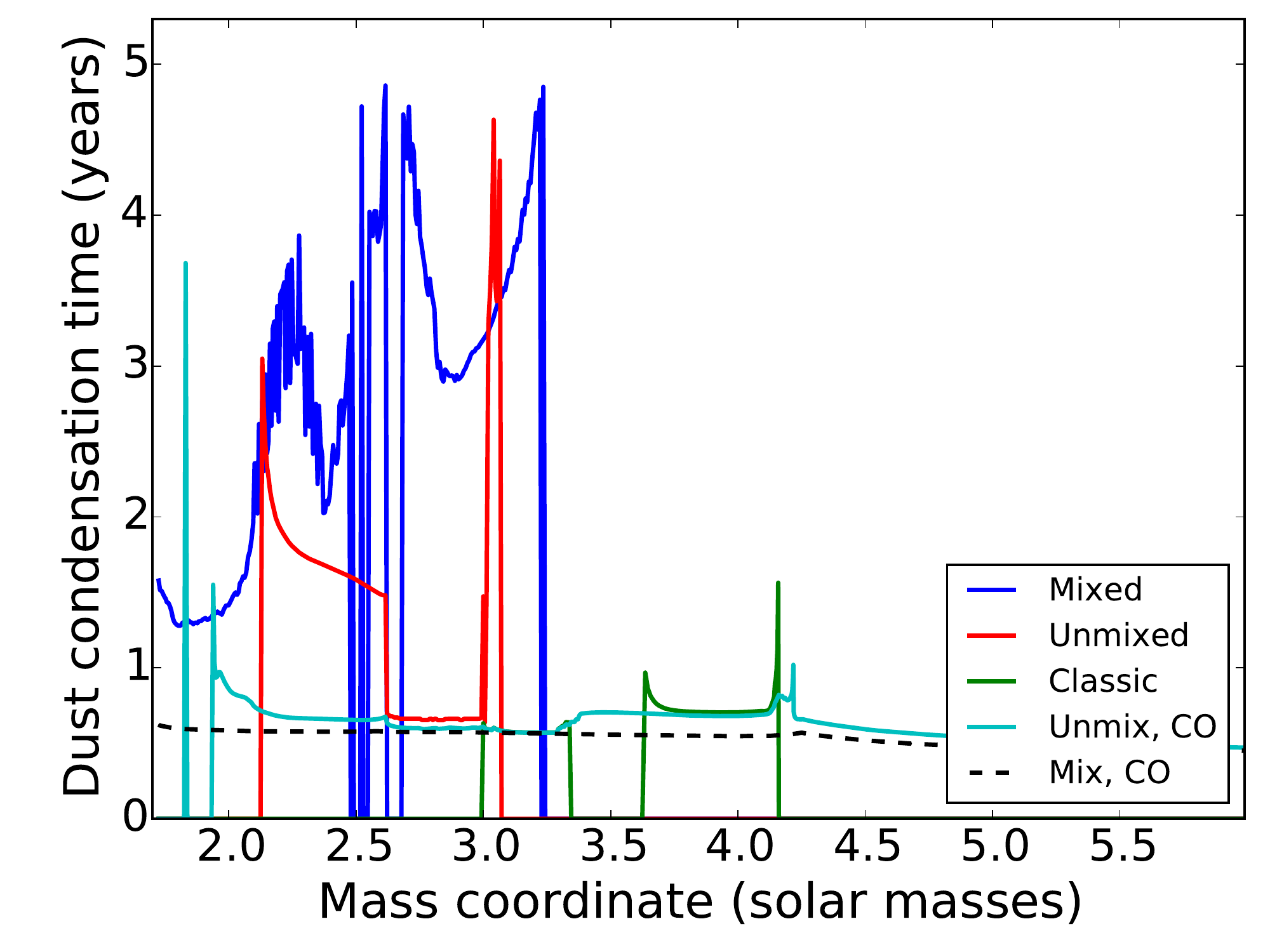}
\caption{Condensation time as a function of the mass coordinate
    (which is monotonically related to the distance from the center of
    the explosion).  The condensation time is defined as the time at
    which $50\,\%$ of the total carbon is condensed into grains.
\label{fig:cond_time}}
\end{figure}

The behavior of the complete models is more complex.  Dust
condensation takes place only in the inner part of the star, within
about $3.5$ solar masses from the center.  In this case, the reason
for the lack of condensation at large distances from the center of the
star is the large number of He cations that erode the carbon clusters
before they can reach a size large enough to become self-shielded. By
the time the radioactivity has ceased and the $\mathrm{He}^+$ fraction
decreases, the gas is too diffuse and cold for efficient nucleation.
Yet, as discussed above, efficiency is large where condensation takes
place, almost always exceeding $50\,\%$, and therefore we do not
expect to see much dust mass growth beyond the five years of the
computation.  Figure~\ref{fig:CO_mass} shows instead the mass of CO in
the entire remnant for the various models (except the classical model,
for which the CO molecule is not considered) as a function of
time. Theoretical predictions are compared against the measurement of
CO in the remnant of SN1987A (Spyromilio et al.\ 1988). Also in this
case the theoretical predictions are in qualitative agreement with the
data, the Mixed model performing better than the unmixed one. Detailed
agreement is not expected since the CCSN progenitor model is not
designed to closely reproduce the progenitor of SN1987A.
Figure~\ref{fig:cond_time} shows the condensation times of the grains
as a function of the mass coordinate.  Here we define the condensation
time as the time at which half of the local carbon is converted into
grains.  The classical models, again, show the lowest diversity, with
dust forming at about seven months at all distances.  The complete
models show instead a distance dependent condensation history.  The
mixed model, owing to it smoother profile of abundances, has a
somewhat monotonic dependence with the earliest formation just outside
of the stellar core, and progressively longer formation times out to a
mass coordinate of $\sim3.2\,\Msun$ where the efficiency drops below
$50\,\%$ and the condensation time is no longer well-defined.  The
unmixed model has instead a more diverse behavior affected by both the
sharp composition changes and the dust formation physics.  Either way,
different condensation times at different radii are the cause of the
more progressive dust formation history seen in
Figure~\ref{fig:dust_mass} for the complete models.

\begin{figure}
\includegraphics[width=\columnwidth]{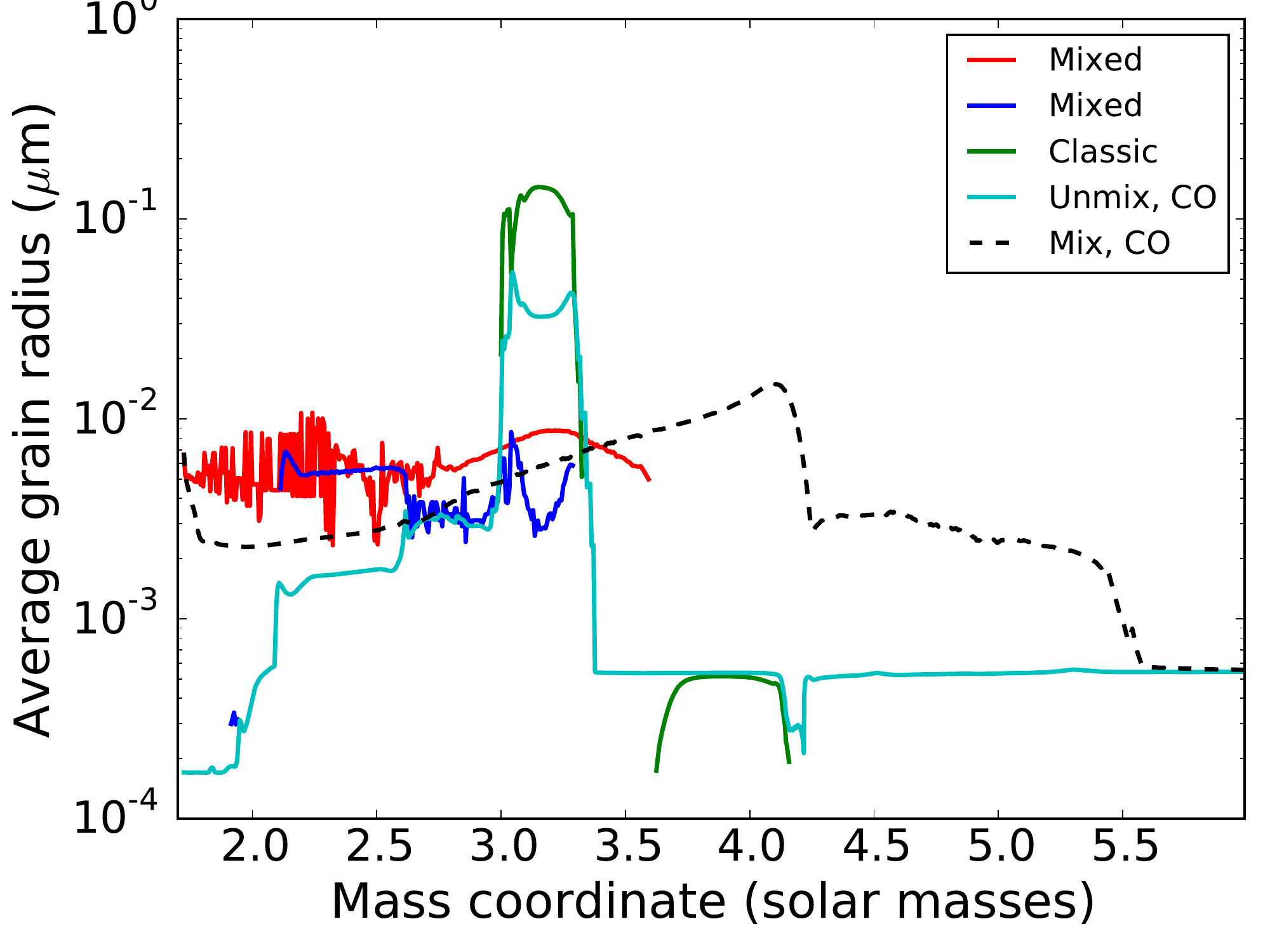}
\caption{Average dust grain size at $5$ years post core collapse as a
    function of the mass coordinate for the various dust condensation
    models.
\label{fig:av_size}}
\end{figure}

Finally, Figure~\ref{fig:av_size} shows the average grain size in each
zone at the end of the calculation.  None of the models has a clearly
understandable pattern, with the exception of the complete mixed model,
for which the biggest grains are formed close to the core.  Grains that
form earlier have more time to grow and therefore are eventually
bigger. The average size, however, depends also on the rate at which
nucleation takes place, and from the speed at which carbon initially
locked into CO molecules is released for use in dust condensation.

\section{Summary and discussion}

We have presented a calculation of carbonaceous dust formation in a
$15\,\Msun$ progenitor star exploding as a core-collapse supernova.
Our study is based on a novel nucleation code that was developed to
address the particular conditions of the exploding supernova.  This is
a rapidly evolving environment with plenty of ionizing radiation and
reactive agents that can harm the forming grains.  Our code uses the
kinetic theory of nucleation in order to be suitable for possible
non-steady state conditions in the fast evolving remnant.  In
addition, the flexibility of the kinetic theory allows us to relax the
capillary approximation for the dimer ($\mathrm{C}_2$) formation and
to include the effect of oxidation and ion-molecule reactions in the
calculation. An alternative framework for joining the chemical and
nucleation approaches has been recently presented by Sarangi \&
Cherchneff (2015). Their approach differs fundamentally from what
presented in this paper in that the chemical phase is joined to a
coalescence phase, rather than to a growth phase. This means that in
their scheme carbon clusters form up to a maximum size of ten atoms
until all the carbon gas has been used and subsequently grow by
coagulation with other clusters rather than grow by adsorption of gas
monomers (or molecules). We find that, at least for carbon, the chance
of collisions of molecular clusters with monomers is far larger than
with other clusters. In our calculations monomers are more abundant
than clusters and, being lighter, have a larger thermal velocity that
makes collisions more frequent. As a subsequent step in understanding
nucleation, however, the coagulation and growth should be integrated
and taken into account simultaneously.

With respect to classical calculations (Kozasa et al.\ 1989, 1991;
Todini \& Ferrara 2001; Nozawa et al.\ 2003; Fallest et al.\ 2011) and
to the results of Sarangi \& Cherchneff (2015), we find that our new
code predicts a much more gradual carbon dust formation, beginning
just a few months after the core collapse and continuing for a few
years.  Despite the more gradual formation, however, we are not able
to fully reproduce the observational results that require an even more
gradual and continuous dust formation in the ejecta, with dust
appearing as early as two months after core collapse and gradually
increasing for a few years to a decade (Gall et al.\ 2014) eventually
leading to a highly efficient condensation of a sizable fraction of a
solar mass (Indebetouw et al.\ 2014). Addition of non-carbonaceous
dust chemistry (e.g., Sarangi \& Cherchneff 2015) and/or a fully
three-dimensional calculation (Lazzati \& Fallest, in preparation) can
ameliorate the discrepancy. It must be also noted that the explosion
model plays a role in the grain formation. A less energetic explosion
than the one we used (or one with more massive ejecta) would result in
higher densities and temperatures at long timescales, allowing for a
longer period of growth and, potentially, more dust condensation. On
the other hand, a more energetic explosion or one with lighter ejecta
would result in colder and less dense gas, likely going earlier into
freeze-out. The role of explosion diversity is certainly as important
as the proper dimensionality of the simulation, since a 3D model is
likely going to contain both faster and slower ejecta than the 1D
spherically simmetric model presented here and elswhere in the
literature.

Even though we consider this work a step forward towards a complete
understanding of dust formation in stellar explosions (and in
general), it is still plagued by some serious limitations.  The most
obvious is the fact that we consider only carbonaceous dust, while
more species are known to condensate in supernova explosions.
Condensation of silicates and other dust species can begin earlier and
explain the early dust formation observed in some SNe (Wooden et al.\
1993; Gall et al.\ 2014; Sarangi \& Cherchneff 2015). On a more
fundamental level, our code still assumes a constant sticking
coefficient $\lambda=1$, a spherical shape for the forming grains,
down to the smallest sizes, and the capillary approximation of a
size-independent surface energy for any cluster with $i>2$.  All these
are important limitations that need to be corrected (see, e.g., Mauney
\& Lazzati 2015) before a serious comparison with data can be
performed.  As any nucleation based work, we also assume that grain
formation and growth takes place by addition of single carbon atoms,
and that grain erosion as well takes place by removing one atom at a
time from a cluster. Reactions that cause the splitting of a cluster
in two fragments are not considered, even though they might be
important, especially at low temperature (see, e.g., Wakelam et al.\
2009). Another important issue is whether the chemical and nucleation
approaches are fully consistent with each other. As we show in
Fig.~\ref{fig:rates}, that might not be the case, creating a serious
issue when they are joined. More theoretical and experimental work
needs to be devoted to understanding the formation rates of carbon
clusters with a few up to a few tens of atoms, and the relative
importance of radiative and non-radiative association processes need
to be pinned down from first principles. The reaction network is also
incomplete, lacking the consideration of the effect of other
potentially harmful cations, such as Ar$^+$ and Ne$^+$. The role of
these cations in silicate nucleation and growth was included in the
work of Sarangi \& Cherchneff (2013, 2015), but ion-molecule reaction
rates for such cations with carbon clusters are not available in the
literature and are therefore omitted in this work.

\acknowledgements We are very grateful to the anonymous referee for
her/his useful and constructive comments. This work was supported in
part by NSF grants AST-1150365 and AST-1461362 (DL), and by an ARC
Future Fellowship FT120100363 (AH).

\end{document}